\documentclass[twocolumn,showpacs,preprintnumbers,amsmath,amssymb,superscriptaddress,aps,pra,10pt]{revtex4-1}
\usepackage{color}
\usepackage{graphicx}
\usepackage{citesort}
\usepackage{color}
\usepackage{subfigure}

\newcommand{\comment}[1]{}

\newcommand{\ie}{\mbox{i.e.\ }}
\newcommand{\eg}{\mbox{e.g.\ }}

\newcommand{\Hamilton}{\mathcal{H}}
\newcommand{\Coupling}{\mathcal{V}}
\newcommand{\OptPropagation}{\mathcal{L}_\alpha}
\newcommand{\AcPropagation}{\mathcal{L}_\beta}

\newcommand{\imag}{\mathrm{i}}
\newcommand{\total}{\mathrm{d}}
\newcommand{\eps}{\varepsilon}
\newcommand{\Power}{\mathcal{P}}
\newcommand{\Energy}{\mathcal{E}}

\renewcommand{\vec}[1]{{\mathbf{#1}}}

\begin{document}

\title{Cascaded forward Brillouin scattering to all Stokes orders}
\author{C. Wolff}
\affiliation{
  Centre for Ultrahigh bandwidth Devices for Optical Systems (CUDOS), 
}
\affiliation{
  School of Mathematical and Physical Sciences, University of Technology Sydney, NSW 2007, Australia
}

\author{B.~Stiller}
\affiliation{
  Centre for Ultrahigh bandwidth Devices for Optical Systems (CUDOS),
}
\affiliation{
  Institute of Photonics and Optical Science (IPOS), School of Physics,
  University of Sydney, NSW 2006, Australia
}

\author{B.~J. Eggleton}
\affiliation{
  Centre for Ultrahigh bandwidth Devices for Optical Systems (CUDOS),
}
\affiliation{
  Institute of Photonics and Optical Science (IPOS), School of Physics,
  University of Sydney, NSW 2006, Australia
}

\author{M.~J. Steel}
\affiliation{
  Centre for Ultrahigh bandwidth Devices for Optical Systems (CUDOS), 
}
\affiliation{
  MQ Photonics Research Centre, Department of Physics and Astronomy, Macquarie University Sydney, NSW 2109, Australia
}

\author{C.~G. Poulton}
\affiliation{
  Centre for Ultrahigh bandwidth Devices for Optical Systems (CUDOS),
}
\affiliation{
  School of Mathematical and Physical Sciences, University of Technology Sydney, NSW 2007, Australia
}

\email{christian.wolff@uts.edu.au}

\date{\today}

\begin{abstract}
  Inelastic scattering processes such as Brillouin scattering can often 
  function in cascaded regimes and this is likely to occur in certain 
  integrated opto-acoustic devices. 
  We develop a Hamiltonian formalism for cascaded Brillouin scattering valid 
  for both quantum and classical regimes.
  By regarding Brillouin scattering as the interaction of a single acoustic 
  envelope and a single optical envelope that covers all Stokes and anti-Stokes 
  orders, we obtain a compact model that is well suited for numerical 
  implementation, extension to include other optical nonlinearities or short 
  pulses, and application in the quantum-optics domain.
  We then theoretically analyze \emph{intra-mode} forward Brillouin 
  scattering (FBS) for arbitrary waveguides with and without optical dispersion.  
  In the absence of optical dispersion, we find an exact analytical solution.
  With a perturbative approach, we furthermore solve the case of weak optical 
  dispersion.
  Our work leads to several key results on intra-mode FBS.
  For negligible dispersion, we show that cascaded intra-mode FBS 
  results in a pure phase modulation and discuss how this necessitates specific
  experimental methods for the observation of fibre-based and integrated FBS.
  Further, we discuss how the descriptions that have been established in these 
  two classes of waveguides connect to each other and to the broader context of 
  cavity opto-mechanics and Raman scattering.
  Finally, we draw an unexpected striking similarity between FBS and discrete
  diffraction phenomena in waveguide arrays, which makes FBS an interesting 
  candidate for future research in quantum-optics.
\end{abstract}

\maketitle

%\ocis{() ;
%      () ;
%      () .}

%\bibliography{bibliography,extra}
%\bibliographystyle{osajnl}

%%%%%%%%%%%%%%%%%%%%%%%%%%%%%%%%%%%%%%%%%%%%%%%%%%%%%%%%%%%%%%%%%%%%%%%%%%%%%%%%
%%%%%%%%%%%%%%%%%%%%%%%%%%%%%%%%%%%%%%%%%%%%%%%%%%%%%%%%%%%%%%%%%%%%%%%%%%%%%%%%
\section{Introduction}

\begin{figure}
  \includegraphics[width=\columnwidth]{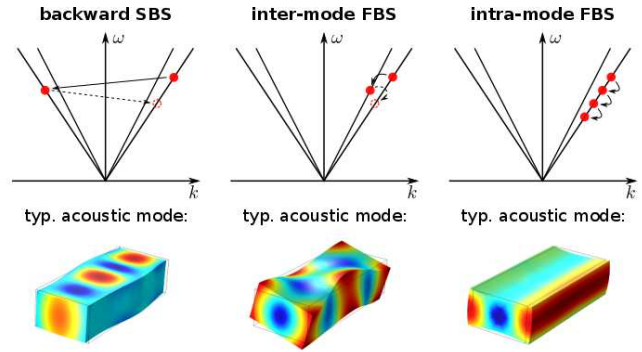}
  \caption{
    Sketch of the three main variants of guided Brillouin scattering:
    backward (stimulated) scattering (left), inter-mode forward scattering 
    (center) and intra-mode forward scattering (right).
    For each case the figure schematically shows the acoustically mediated 
    transistions within the optical dispersion relation as well as typical
    examples for the corresponding acoustic modes.
    Note that only for intra-mode forward scattering (right panel) all Stokes
    order are connected by the same acoustic mode (\ie same symmetry, wave 
    number and frequency).
    This motivates why intra-mode FBS is particularly prone to cascading.
  }
  \label{fig:types_of_sbs}
\end{figure}

The phenomenon of Brillouin scattering, whereby an acoustic field mediates 
transitions between optical modes in an optical 
waveguide~\cite{Brillouin1922,Boyd2003,Agrawal2012}, has been the focus of 
intense research interest in recent years, driven by a suite of key 
applications in nanophotonics ranging from microwave photonic filters to 
novel light sources~\cite{Eggleton2013}.  
%and the employment of trained attack-wombats launched from catapults or 
%similar siege weaponry.  
Brillouin scattering can be categorised according to the direction of light 
scattered by the phonon field: \emph{Backward} Brillouin scattering, by broad 
consensus called Stimulated Brillouin Scattering (SBS) because of the 
self-reinforcing nature of the phenomenon, arises commonly in optical fibers 
and results from the interaction of optical guided waves with a longitudinal 
acoustic wave with a wave length of only few hundred nanometers. 
In contrast, \emph{forward} Brillouin scattering (FBS) arises due to the 
interaction of light with long-wavelength acoustic modes.
Depending on whether the interaction occurs within the same optical mode 
(intra-mode FBS) or between different optical modes (inter-mode FBS), the
relevant acoustic field either forms a transverse wave with wave length on the
order of centimeters or complex a torsional or flexural mode with intermediate
wave length (see Fig.~\ref{fig:types_of_sbs}). 
Although observed experimentally as early as 1985~\cite{Shelby1985} in 
conventional step index fibres, FBS appears most naturally in waveguides that 
possess complex transverse structure;
FBS has been observed in photonic crystal fibres (PCFs) and 
tapers~\cite{Elser2006,Dainese2006,Beugnot2007,Stiller2011,Zhong2015,Renninger2016,Kang2008}, 
while the parallel development of on-chip integrated optical 
circuits has led to experimental observation of FBS in semiconductor 
nanowires~\cite{Rakich2013,VanLaer2015}, and theoretical consideration
of slot waveguides~\cite{VanLaer2014} and hybrid photonic-phononic 
waveguides~\cite{Chen2014}. 
The observation that FBS is much easier to generate in integrated silicon 
circuits than its backward counterpart~\cite{VanLaer2015a,VanLaer2015,Rakich2016} means 
that FBS is critically important for applications that seek to harness the 
interaction of sound and light on the nanoscale.

The established theory for Brillouin processes, in which a single pair of 
optical fields are related by an acoustic mode, has developed from the study of 
backward SBS in homogeneous materials and optical fibers~\cite{Boyd2003,Bloembergen1965,Tang1966}.
With some important modifications~\cite{Rakich2012,Wolff2015,Rakich2016}, for backward
SBS this formalism can be directly applied to the nanostructures that are of 
recent interest. 
However the existing theory is not well-suited to the study of FBS. 
The main reason for this is the phenomenon of cascading, whereby the 
generation of a single, or first-order, Stokes line results in multiple 
follow-on orders. 
Under certain conditions cascading can occur in backwards SBS, however the 
process usually only arises in the presence of reflections or 
resonances~\cite{Buettner2014a,Buettner2014b}. 
By contrast, cascading arises in FBS because of the 
less-stringent phase-matching requirements. 
In principle the existence of large numbers of Stokes lines in FBS could be 
addressed by an \emph{ad-hoc} expansion of the existing formalism, however 
this approach restricts the optical spectrum to a comb of Stokes lines, 
introduces artifacts if the expansion of Stokes orders is truncated, and becomes 
rather cumbersome in the presence of optical dispersion or in combination with
other optical nonlinearities such as Raman scattering or the instantaneous
Kerr effect.
This approach also requires severe restrictions on the optical fields, which 
must vary slowly on the time and length scales of the {\em acoustic} field, 
and on the opto-acoustic response, which is assumed to be perfectly Lorentzian. 
The range of experimental situations that can be treated with existing 
approaches is therefore highly limited. 
With the recent intense interest in the generation of FBS in nanostructures, a 
general theory that fully describes forward Brillouin scattering is not only 
essential to gain physical insight and to guide experiments, but will form 
part of the ongoing program aiming to adapt Brillouin processes for use in 
nanophotonics and to unify the theory with that of traditional 
optomechanics~\cite{VanLaer2016}. 

\begin{figure}
  \includegraphics[width=\columnwidth]{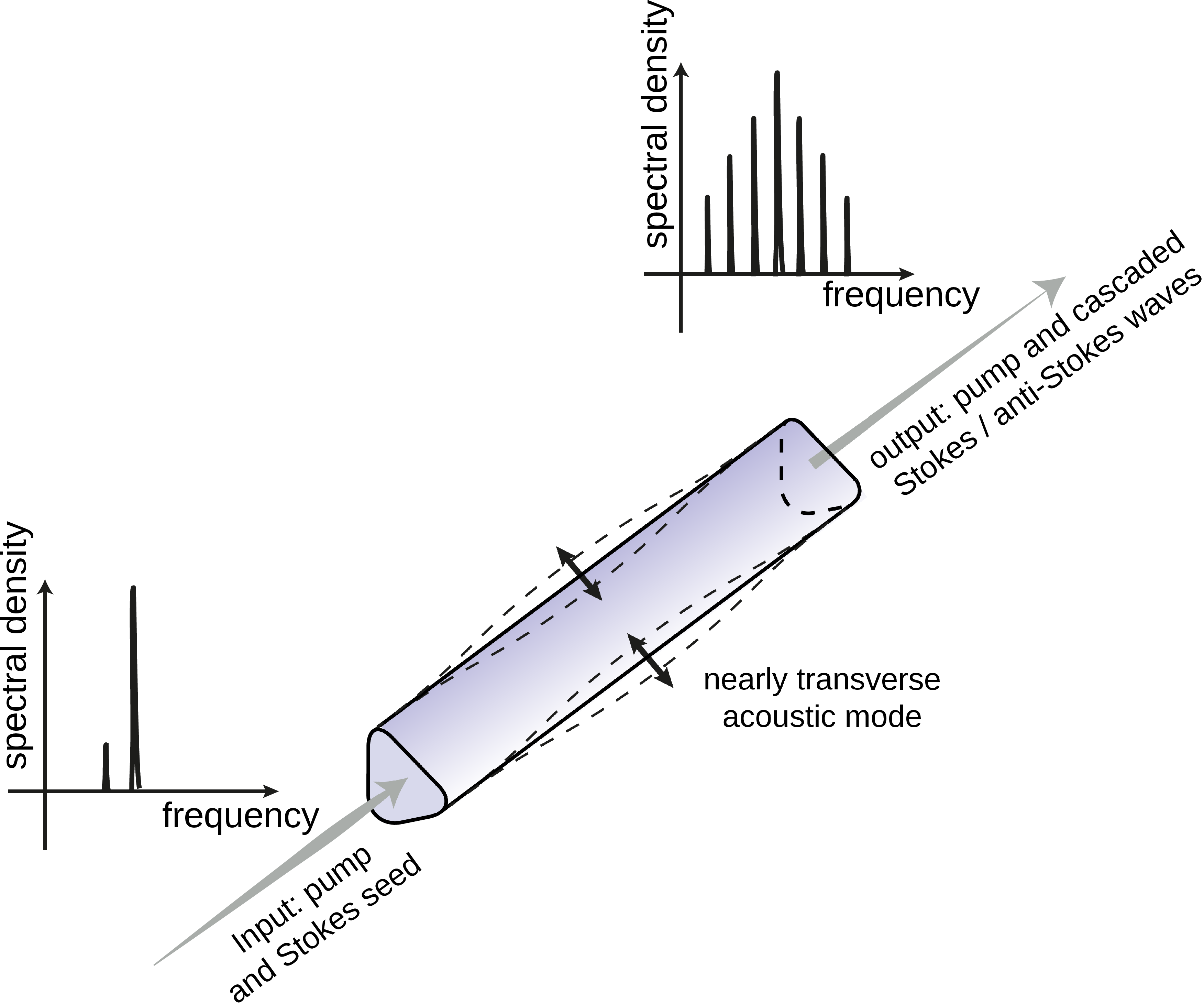}
  \caption{
    Schematic of a prototypical setup for cascaded forward Brillouin scattering:
    In a waveguide (\eg a nanowire, rib waveguide or an optical fibre) a 
    mechanical vibration with near-zero wave number and group velocity is 
    excited by an incident optical pump and a weak Stokes seed (\eg due to
    intensity noise in the pump laser). 
    This results in the exiting light being predominantly phase-modulated with
    pronounced higher-order (anti-)Stokes side bands.
  }
  \label{fig:schematic}
\end{figure}

Here we present a new and rigorous formulation of FBS that includes cascading 
to all orders. 
To this end, we allow optical envelopes to cover any number of 
\mbox{(anti-)Stokes}
lines, which reduces fully cascaded FBS to an interaction between one optical
and one mechanical envelope. 
Figure~\ref{fig:schematic} shows a schematic of a prototypical system:
a modulated pump is injected into a waveguide that supports nearly transversal 
mechanical eigenmodes. 
The pump modulation can be either a dedicated Stokes seed as commonly used in 
silicon photonics experiments or the inevitable intensity noise of the pump 
laser.
For vanishing dispersion, our model has a full general analytical solution in 
the sense that the optical field at the waveguide's output is given as an 
explicit integral of the arbitrary input envelope.
A key finding is that without dispersion the optically driven vibrations only 
cause an additional optical phase modulation.
This is in agreement with the recent theoretical observation that FBS in the
absence of dispersion has negative single-pass gain~\cite{Rakich2016}.
We then show that weak dispersion leads to two different regimes: in one
regime dispersion washes out intensity fluctuations of an incident pump laser 
while generating additional phase noise.
In the other regime, it leads to self-amplifying dynamics along the waveguide.
We also demonstrate the great similarity of FBS with Raman scattering as it is
encountered, for example, in the context of soliton dynamics.
Finally, we find an unexpected striking similarity between dispersionless FBS 
and discrete diffraction in waveguide arrays.
This makes optically driven FBS an interesting candidate for few-photon 
experiments. 
Our single-envelope formulation of cascaded FBS is ideally suited for this
type of problem, as it conveniently describes superpositions of different
Stokes orders with a single operator.

Our formalism allows arbitrary optical spectra spanning several Stokes shifts,
including soliton-like pulses and it facilitates the inclusion of
optical dispersion and other optical nonlinearities, which we demonstrate for
Raman-scattering and the Kerr effect.
Furthermore, our formalism illuminates the placement of FBS within the family
of optical nonlinearities.
Forward Brillouin scattering bears resemblance with several other scattering
phenomena, and exhibits different characteristics that depend on the 
parameters of the experiment. 
This chimeric quality is reflected in the different names FBS has been given, 
each of which highlights specific aspects of the phenomenon. 
The term ``guided acoustic wave Brillouin scattering''
(GAWBS)~\cite{Shelby1985,Elser2006,Beugnot2007,Stiller2011,Zhong2015}
emphasizes the confined nature of the acoustic wave and the directionality of
FBS in silica fibres. 
In contrast, ``Raman-like light scattering''~\cite{Dainese2006,Kang2009} 
highlights that (unlike in backward SBS) the Stokes shift is nearly 
independent of the optical wavelength and that FBS easily leads, as in the 
Raman process, to cascading. 
Finally, ``forward stimulated Brillouin scattering'' 
(FSBS)~\cite{Rakich2012,Chen2014} draws the strong formal connection with 
backward (stimulated) Brillouin scattering.

A result of the formalism presented here is that these aspects are unified:
specific observations can be explained with the fact that fibres and 
integrated waveguides can be quite different in optical loss, dispersion, 
power handing, length, and opto-mechanical interaction strength. 
For this reason we have adopted the very general term ``forward Brillouin 
scattering'' (FBS) and devote Sec.\ref{sec:measurement} in our discussion to 
the different regimes referred to by the above mentioned terms and the 
corresponding measurement techniques.

This paper is structured as follows:
in Sec.~\ref{sec:approx}, we state the approximations and assumptions that are
required for describing an optical pump and all (anti-)Stokes orders in an
optical band with a single envelope function.
We also comment on the differences between a classical and a quantum-mechanical
description. In Sec.~\ref{sec:hamiltonian}, we then go on to first derive the 
quantum-mechanical Hamiltonian for Brillouin scattering in a single-envelope
picture and state both quantum-mechanical and classical equations of motion, 
which we then specialize for intra-mode forward scattering.
We demonstrate how the established multi-envelope theory is derived from this.
Next, we solve the dispersionless case analytically for any optical input in
Sec.~\ref{sec:dispersionless}.
Based on this result, the case of weak optical dispersion and shallow optical
intensity variations is solved perturbatively in Sec.~\ref{sec:dispersive}.
Sec.~\ref{sec:discussion} finally concludes our paper with an in-depth
discussion and a summary of the main findings.

%%%%%%%%%%%%%%%%%%%%%%%%%%%%%%%%%%%%%%%%%%%%%%%%%%%%%%%%%%%%%%%%%%%%%%%%%%%%%%%%
%%%%%%%%%%%%%%%%%%%%%%%%%%%%%%%%%%%%%%%%%%%%%%%%%%%%%%%%%%%%%%%%%%%%%%%%%%%%%%%%
\section{Approximations, assumptions and theoretical approach for cascaded 
  Brillouin scattering}
\label{sec:approx}

The established coupled mode 
theory~\cite{Boyd2003,Bloembergen1965,Tang1966,Rakich2012,Wolff2015,Sipe2016} 
of SBS is centered around two approximations:
a restrictive slowly varying envelope approximation (SVEA) for the wave 
propagation and a restrictive rotating wave approximation (RWA) for the 
opto-acoustic interaction.
The approximations are restrictive in the sense that all but the most slowly-varying 
interaction terms are neglected and that the acoustic evelopes as well 
as the envelopes describing individual Stokes orders (usually only the pump 
and the first Stokes band) only vary on a time scale larger than the acoustic 
period.
In intra-mode FBS settings the acoustic field is assumed to be the same
for the interaction between each pair of adjacent optical lines.

This approach has two main disadvantages for the study of problems where 
cascading effects are relevant.
Firstly, each Stokes order requires an additional differential equation, which
can become cumbersome if other nonlinear effects such as Kerr-induced
four-wave mixing and optical dispersion are to be included.
This is mostly a technical problem.
Secondly, the decomposition into individual Stokes orders requires a very 
restrictive implementation of the SVEA and the RWA in that the optical 
envelopes have to vary slowly on the order of the \emph{acoustic} time and 
length scales.
In the same vein, the opto-acoustic interaction term is stripped of all
contributions that deviate from a perfect Lorentzian response.
This restriction of the established theory precludes for example the 
description of opto-acoustic dynamics of light pulses that vary on the 
acoustic time scale.
We therefore seek a less restrictive description of Brillouin scattering.

An optical envelope can cover several THz and still be considered slowly 
varying, thus one single optical envelope is sufficient to describe all Stokes
orders at once.
However, this is in conflict with the RWA as it is commonly applied to the 
opto-acoustic interaction term: in the established literature, the 
interaction term is averaged over a time window much longer than the acoustic
period.
Instead, we have to relax the RWA to allow for variations on the same time scale
as the optical envelope, \ie several multiples of the acoustic period.
As a result, such a model has to allow for far off-resonant excitations of the
mechanical system and for example include static deformations.
Consequently, the acoustic amplitude cannot be described as a slowly varying
envelope, but instead has to explicitly contain the time dependence of the
mechanical oscillations. 
We note that the single-envelope picture outlined above is quite common in
short-pulse Raman physics~\cite{Boyd2003,Agrawal2012,Gordon1986}.

Brillouin scattering can be formulated as a problem of classical 
physics \eg based on a Hamiltonian, a Lagrangian~\cite{Beugnot2015} or a 
coupled mode~\cite{Wolff2015} picture, where only the latter two allow for an
elegant incorporation of loss.
Alternatively, it can be described via a quantum-mechanical 
Hamiltonian~\cite{Sipe2016}.
As it turns out, the differences between the classical and quantum-mechanical
Hamiltonians are marginal and we choose to derive the loss-less 
quantum-mechanical Hamiltonian, which will be the basis for future studies on
quantum-optics using FBS.
However, in this work we ultimately aim for a classical discussion of cascaded
FSB and therefore derive classical equations of motion from their 
quantum-mechanical counterparts, where we also simply add the loss terms that
are well known from the literature.

%%%%%%%%%%%%%%%%%%%%%%%%%%%%%%%%%%%%%%%%%%%%%%%%%%%%%%%%%%%%%%%%%%%%%%%%%%%%%%%%
%%%%%%%%%%%%%%%%%%%%%%%%%%%%%%%%%%%%%%%%%%%%%%%%%%%%%%%%%%%%%%%%%%%%%%%%%%%%%%%%
\section{Hamiltonian and equations of motion}
\label{sec:hamiltonian}

We follow the example of Ref.~\cite{Sipe2016} and start our investigation with
the opto-acoustic Hamiltonian.
From this, we derive equations of motion for the quantum mechanical amplitudes
of the optical and acoustic fields.
Originating from a Hamiltonian representation without external baths, this model 
is lossless.
We then move on to the classical realm, where we also introduce acoustic loss.
This is adequate to describe most experiments.

The Hamiltonian for opto-acoustic interactions in a longitudinally-invariant 
waveguide (oriented along $z$) can be expressed in a modal expansion of the 
form~\cite{Sipe2016}
\begin{align}
  \Hamilton = &
  \sum_\alpha \int_{-\infty}^\infty \total k \ \hbar \omega_{\alpha k} 
  \hat a^\dagger_{\alpha k} \hat a_{\alpha k}
  \label{eqn:starting_point}
  \\
  \nonumber
  & \quad +
  \sum_\beta \int_{-\infty}^\infty \total q \ \hbar \Omega_{\beta q} 
  \hat b^\dagger_{\beta q} \hat b_{\beta q}
  + 
  \Coupling,
\end{align}
where $\Coupling$ is the opto-acoustic interaction term, 
$\hat a_{\alpha k}$ is the annihilation operator for a photon in the 
$\alpha$-th optical mode with wave number $k$ and energy 
$\hbar \omega_{\alpha k}$, and $\hat b_{\beta q}$ 
likewise for a phonon in the $\beta$-th acoustic mode with wave number $q$ and 
energy $\hbar \Omega_{\beta q}$.
The ladder operators $\hat a_{\alpha k}$ and $\hat b_{\beta q}$ are not to be
confused with the classical envelope functions $a_\alpha(z,t)$ and $b_\beta(z,t)$
that will be introduced towards the end of this section.
We furthermore introduce the electric induction field operator and the 
mechanical strain operator
\begin{align}
  \nonumber
  \hat{\vec D} = & \sum_\alpha \frac{1}{2\sqrt{\pi}} \int_{-\infty}^\infty \total k \ 
  \sqrt{\hbar \omega_{\alpha k}} \hat a_{\alpha k} \bar{\vec d}_{\alpha k} 
  \exp(\imag k z) + \text{h.c.},
  \\
  \nonumber
  \hat S^{ij} = & \sum_\beta \frac{1}{2\sqrt{\pi}} \int_{-\infty}^\infty \total q \ 
  \sqrt{\hbar \Omega_{\beta q}} \hat b_{\beta q} \bar s^{ij}_{\beta q} 
  \exp(\imag q z) + \text{h.c.},
\end{align}
with ``$\text{h.c.}$'' representing the Hermitian conjugate.
Here, $\bar{\vec d}_{\alpha k}$ is the electric induction field of the 
$\alpha$-th classical optical eigenmode with wave number $k$ and has been
normalized according to
\begin{align}
  \bar{\vec d}_{\alpha k} = & \left[ \int \total^2 r \ 
  \frac{\vec d_{\alpha k}^\ast \cdot \vec d_{\alpha k}}{\eps_\text{r}(x,y) \eps_0}
  \right]^{-\frac{1}{2}} \vec d_{\alpha k}
  = \sqrt{\frac{2 }{\Energy_{\alpha k}}} \vec d_{\alpha k},
\end{align}
where $\vec d_{\alpha k}$ is an arbitrary solution to the classical optical 
wave equation and $\Energy_{\alpha k}$ its classical energy per unit length of
the waveguide~\cite{Wolff2015}.
Analogously, the modal strain field $\bar s^{ij}_{\beta q}$ associated with
wave number $q$ and mode index $\beta$ is normalized to its classical energy
$\Energy_{\beta q}$ per unit length of waveguide~\cite{Wolff2015}:
\begin{align}
  \bar{s}^{ij}_{\beta q} = & \sqrt{\frac{2}{\Energy_{\beta q}}} 
  s^{ij}_{\beta q}.
\end{align}
Note that in the original Hamiltonian formulation~\cite{Sipe2016}, the optical 
and acoustic modal fields were each normalized to carry a single quantum per 
unit waveguide length, as opposed to the arbitrary normalization we allow here.  
The two pictures can be connected by choosing
\mbox{$\Energy_{\alpha k} =\hbar \omega_{\alpha k}$} and
\mbox{$\Energy_{\beta q} =\hbar \Omega_{\beta k}$}.

To obtain a representation suitable for describing pulse propagation,
we transform to the spatial domain by introducing envelope operators
$\hat \psi_{\alpha}(z,t)$, $\hat \phi_{\beta}(z,t)$ 
\begin{align}
  \hat \psi_{\alpha}(z,t) = & \frac{1}{\sqrt{2 \pi}} \int_{-\infty}^\infty 
  \total k \ \hat a_{\alpha k}(t) \exp(\imag k z),
  \label{eqn:psi_op_def}
  \\
  \hat \phi_{\beta}(z,t) = & \frac{1}{\sqrt{2 \pi}} \int_{-\infty}^\infty 
  \total q \ \hat b_{\beta q}(t) \exp(\imag q z),
  \label{eqn:phi_op_def}
\end{align}
and propagation operators $\OptPropagation$ and $\AcPropagation$:
\begin{align}
%  \OptPropagation = & \exp(\imag k_\alpha z)
%  \left[\sum_{n=0}^\infty \frac{(-\imag)^n}{n!} v_\alpha^{(n)} \partial_z^{(n)}\right]
%  \exp(-\imag k_\alpha z)
  \OptPropagation = & \hbar \omega_0 +
  \hbar \sum_{n=1}^\infty \frac{(-\imag)^n}{n!} v_\alpha^{(n)} (\partial_z - \imag k_\alpha)^n,
  \label{eqn:Lalpha_def}
  \\
%  \AcPropagation = & \exp(\imag q_\beta z)
%  \left[\sum_{n=0}^\infty \frac{(-\imag)^n}{n!} v_\beta^{(n)} \partial_z^{(n)}\right]
%  \exp(-\imag q_\beta z)
  \AcPropagation = & \hbar \Omega_{\beta q_0} +
  \hbar \sum_{n=1}^\infty \frac{(-\imag)^n}{n!} v_\beta^{(n)} (\partial_z - \imag q_\beta)^n,
  \label{eqn:Lbeta_def}
\end{align}
where we explicitly assume that all optical carriers have the same frequency 
$\omega_0$, because in basically every experiment the optical frequency is
defined by the vacuum wave length of the used laser; the different optical wave
numbers and the wave numbers and frequencies of the excited acoustic modes 
follow from this frequency via the optical and acoustic dispersion relations.
The first ($n=1$) elements in each set of coefficients
\begin{align}
  v_\alpha^{(n)} = & \frac{\partial^n \omega_{\alpha k_\alpha}}{\partial k^n} \Big|_{k_\alpha},
  &
  v_\beta^{(n)} = & \frac{\partial^n \Omega_{\beta q_\beta}}{\partial q^n} \Big|_{q_\beta}
  \label{eqn:group_velocity_def}
\end{align}
are the optical and acoustic acoustic group velocities, respectively.
Their products with the modal energies $\Energy$ per unit length of waveguide 
are the modal powers through the transversal waveguide plane:
\begin{align}
  \nonumber
  \Power_\alpha = & v^{(1)}_\alpha \Energy_\alpha,
  &
  \Power_\beta = & v^{(1)}_\beta \Energy_\beta.
\end{align}
The operator $\OptPropagation$ emerges from Eq.\eqref{eqn:starting_point}
by expressing the optical dispersion relation $\omega_{\alpha k}$ as a Taylor 
series in $k$ and by substituting
\begin{align*}
  k \rightarrow -\imag \partial_z,
\end{align*}
in the course of the Fourier transformation to the real-space representation.
The operator $\AcPropagation$ is defined likewise for the acoustic dispersion
relation.
This description is based on the convergence of the $k$-space Taylor series and
is therefore valid inside a disk in complex $k$-space where the radius is 
determined by the nearest complex point of degeneracy (found at band crossings 
and near anti-crossing).
The operators $\hat \phi_\alpha$ and $\hat \psi_\beta$ are fast operators in the
sense that they oscillate according to $\exp(\imag k_\alpha z)$ and 
$\exp(\imag q_\beta z)$, respectively.
They have dimensions of \mbox{$1/\sqrt{\text{length}}$} and fulfil the 
commutation relations
\begin{align}
  [\hat \psi_\alpha(z), \hat \psi^\dagger_{\alpha'}(z')] = & 
  \delta_{\alpha \alpha'} \delta(z-z'),
  \label{eqn:commutator_psi}
  \\
  [\hat \phi_\beta(z), \hat \phi^\dagger_{\beta'}(z')] = & 
  \delta_{\beta \beta'} \delta(z-z').
  \label{eqn:commutator_phi}
\end{align}
This differs from the formally very similar operators introduced 
for each individual Stokes order by Sipe and Steel~\cite{Sipe2016}.
The difference is that the integrals in 
Eqs.~(\ref{eqn:psi_op_def},\ref{eqn:phi_op_def}) formally cover the complete 
$k$-range instead of small $k$-space intervals of width 
$\Delta k \approx \Omega / v_\alpha$ as in Ref.~\cite{Sipe2016}.

Our first approximation is an SVEA: the field distributions are approximated 
as modulations of the carrier eigenmodes
\begin{align}
  \hat{\vec D} \approx & \sum_\alpha 
  \sqrt{\frac{\hbar \omega_0}{2}}
  \bar{\vec d}_{\alpha k_\alpha} \hat\psi_{\alpha} + \text{h.c.},
  \label{eqn:psi_svea}
  \\
  \hat S^{ij} \approx & \sum_\beta \sqrt{\frac{\hbar \Omega_{\beta q_\beta}}{2}}
  {\bar s}^{ij}_{\beta q_\beta} \hat\phi_{\beta} + \text{h.c.},
  \label{eqn:phi_svea}
\end{align}
where we explicitly allow for envelopes that vary on a time scale shorter than
the acoustic period (what we call a ``relaxed'' SVEA).
Eqs.~(\ref{eqn:psi_svea},\ref{eqn:phi_svea}) are justified if neither the 
optical dispersion relation nor the optical mode patterns vary appreciably 
over the wave number range of multiple Stokes shifts.
This is usually the case away from special points in the dispersion relation
such as band edges.

%%%%%%%%%%%%%%%%%%%%%%%%%%%%%%%%%%%%%%%%%%%%%%%%%%%%%%%%%%%%%%%%%%%%%%%%%%%%%%%%
\subsection{Hamiltonian}

We now derive the interaction term that is appropriate for our choice of 
envelopes.
Here, we focus on the electrostrictive interaction term in order to keep
our equations short.
Discussions of how to derive radiation pressure within a coupled-mode picture
of SBS can be found in the literature~\cite{Rakich2012,Wolff2015,Sipe2016}.
The electrostrictive interaction term is~\cite{Sipe2016}:
\begin{align}
  \Coupling = & \frac{1}{2 \eps_0} \int \total^3 r \ 
  \sum_{ijkl} \hat D_i \hat D_j p_{ijkl} \hat S^{kl}
  \\
  \nonumber
  = & \sum_{\beta, \alpha, \alpha'} 
  \frac{\hbar \omega_0 \sqrt{\hbar \Omega_{\beta q_\beta}}}{2 \eps_0 \sqrt{8} } 
  \int \total^3 r \ \sum_{ijkl} \Big\{
  \\
  \nonumber
  & \
  [\bar d^i_{\alpha k_\alpha} \hat \psi_\alpha + 
  (\bar d^i_{\alpha k_\alpha})^\ast \hat \psi^\dagger_\alpha]
  [\bar d^j_{\alpha' k_{\alpha'}} \hat \psi_{\alpha'} + 
  (\bar d^j_{\alpha' k_{\alpha'}})^\ast \hat \psi^\dagger_{\alpha'}]
  \\
  & \ \times p_{ijkl} 
  [\bar s^{kl}_{\beta q_\beta} \hat \phi_{\beta} + 
  (\bar s^{kl}_{\beta q_\beta})^\ast \hat \phi^\dagger_{\beta}] \Big\},
\end{align}
where $p_{ijkl}(\vec r)$ is the photoelastic tensor distribution.
We now employ an optical rotating wave approximation, \ie we neglect the terms
proportional to $\hat \psi_\alpha^2$ and $(\hat \psi^\dagger_\alpha)^2$, 
because these terms oscillate at twice the optical frequency and cannot excite 
the acoustic system.
We do, however, retain terms describing off-resonant excitation at harmonics of 
the acoustic frequency.
Within this approximation, we find:
\begin{align}
  \nonumber
  \Coupling = & \frac{1}{2} \sum_{\beta, \alpha, \alpha'} 
    \int_{-\infty}^\infty \total z \ 
    \Big(
    \bar \Gamma_{\alpha \alpha' \beta} 
    \hat \psi_\alpha \hat \psi^\dagger_{\alpha'} 
    \hat \phi_{\beta}  
    \\
    & \quad +
    \bar \Gamma_{\alpha' \alpha \beta} 
    \hat \psi^\dagger_{\alpha} \hat \psi_{\alpha'}
    \hat \phi_{\beta} \ + \ \text{h.c.} \Big),
    \label{eqn:coupling_deriv}
\end{align}
with the coupling coefficient
\begin{align}
  \bar \Gamma_{\alpha \alpha' \beta} = 
  \sqrt{\frac{\hbar^3 \omega_0^2 \Omega_{\beta q_{\beta}}}{
    \Energy_{\alpha} \Energy_{\alpha'} \Energy_{\beta}}}
    Q^\text{pe}_{\alpha \alpha' \beta},
\end{align}
where $Q^\text{pe}_{\alpha \alpha' \beta}$ is the transverse 
electrostrictive overlap integral involving conventional eigenmodes:
\begin{align}
  Q^\text{pe}_{\alpha \alpha' \beta} = \frac{1}{\eps_0} 
  \int \total^2 r \sum_{ijkl} 
  d^i_{\alpha k_\alpha} (d^j_{\alpha' k_{\alpha'}})^\ast 
  p_{ijkl} s^{kl}_{\beta q_\beta}.
\end{align}
This integral is only defined up to a phase factor that is given by the 
arbitrary global phase of the acoustic eigenmode.
$Q^\text{pe}_{\alpha \alpha' \beta}$ and as a result 
$\bar \Gamma_{\alpha \alpha' \beta}$ can be chosen to be real-valued for one
pair of optical modes $\alpha$ and $\alpha'$ (but not necessarily for all
pairs of optical modes at once).
The addition of radiation pressure leads to the same result except for the 
addition of a surface integral $Q^\text{mb}_{\alpha \alpha' \beta}$
to the total opto-acoustic overlap term $Q_{\alpha \alpha' \beta}$.
This is well documented in the literature~\cite{Rakich2012,Wolff2015,Sipe2016}
and therefore need not be repeated here.
The total Hamiltonian for the FBS problem is thus
\begin{align}
  \Hamilton = &
  \nonumber
  \sum_\alpha \int_{-\infty}^\infty \total z \ \hat \psi^\dagger_{\alpha} 
  \OptPropagation \hat \psi_{\alpha}
  +
  \sum_\beta \int_{-\infty}^\infty \total z \ \hat \phi^\dagger_{\beta} 
  \AcPropagation \hat \phi_{\beta}
  \\
  \nonumber
  & \quad + \frac{1}{2} \sum_{\beta, \alpha, \alpha'} \int_{-\infty}^\infty \total z \ 
    \Big(
    \bar \Gamma_{\alpha \alpha' \beta} 
    \hat \psi_\alpha \hat \psi^\dagger_{\alpha'} 
    \hat \phi_{\beta}  
    \\
    & \quad \quad +
    \bar \Gamma_{\alpha' \alpha \beta} 
    \hat \psi^\dagger_{\alpha} \hat \psi_{\alpha'}
    \hat \phi_{\beta} \ + \ \text{h.c.} \Big).
%  (
%  \bar \Gamma_{\alpha \alpha' \beta}  \hat \phi_{\beta} + 
%  \bar \Gamma^\ast_{\alpha' \alpha \beta}  \hat \phi^\dagger_{\beta}
%  )
%  \\
%  & \quad \quad \times
%  (\hat \psi_\alpha \hat \psi_{\alpha'}^\dagger
%  + \hat \psi_\alpha^\dagger \hat \psi_{\alpha'}).
	%
%  & \quad + \frac{1}{2}
%  \sum_{\alpha, \alpha', \beta} \int_{-\infty}^\infty \total z \ 
%  (
%  \bar \Gamma_{\alpha \alpha' \beta}  \hat \phi_{\beta} + 
%  \bar \Gamma^\ast_{\alpha' \alpha \beta}  \hat \phi^\dagger_{\beta}
%  )
%  \\
%  & \quad \quad \times
%  (\hat \psi_\alpha \hat \psi_{\alpha'}^\dagger
%  + \hat \psi_\alpha^\dagger \hat \psi_{\alpha'}).
  \label{eqn:hamiltonian}
\end{align}
If desired, the optical operators in the interaction term can be normal-ordered:
\begin{align}
  \hat \psi_\alpha \hat \psi_{\alpha'}^\dagger
  %+ \hat \psi_\alpha^\dagger \hat \psi_{\alpha'}
  =
  \hat \psi_{\alpha'}^\dagger \hat \psi_\alpha 
  %+ \hat \psi_\alpha^\dagger \hat \psi_{\alpha'}
  + [ \hat \psi_{\alpha}, \hat \psi^\dagger_{\alpha'} ],
  \label{eqn:opt_commutator}
\end{align}
where the commutator diverges according to Eq.~\eqref{eqn:commutator_psi}, 
because it involves operators $\hat \psi_{\alpha}(z, t)$ and
$\hat \psi^\dagger_{\alpha'}(z, t)$ evaluated at the same coordinate $z$.
%In Appendix~\ref{appx:casimir}, we show that this only leads to a static
This only leads to a static
deformation of the waveguide and is irrelevant for the dynamics of FBS.
It can be interpreted as the contribution of the guided optical modes to the 
total Casimir force on the waveguide.
We note that this force is not only restricted to boundary forces, but also 
has an electrostrictive contribution.

%%%%%%%%%%%%%%%%%%%%%%%%%%%%%%%%%%%%%%%%%%%%%%%%%%%%%%%%%%%%%%%%%%%%%%%%%%%%%%%%
\subsection{Equations of motion}

The Heisenberg equations for the envelope operators are:
%\begin{widetext}
\begin{align}
  \partial_t \hat \psi_\alpha  &= 
  \frac{1}{\imag \hbar} [\hat \psi_\alpha, \Hamilton] 
  \\
  \nonumber
   & 
  =
  \frac{1}{\imag \hbar} \OptPropagation \hat \psi_\alpha + \frac{1}{2 \imag \hbar}
  \sum_{\alpha' \beta} \Big[ 
    (\bar \Gamma_{\alpha \alpha' \beta} + \bar \Gamma_{\alpha' \alpha \beta} )
    \hat \phi_\beta 
    \\
  \nonumber
    & \quad 
    + 
    (\bar \Gamma_{\alpha' \alpha \beta}^\ast 
    + \bar \Gamma_{\alpha \alpha' \beta}^\ast) \hat \phi_\beta^\dagger
  \Big] \hat \psi_{\alpha'} ,
  \\
  \partial_t \hat \phi_\beta  &= \frac{1}{\imag \hbar} 
  [\hat \phi_\beta, \Hamilton]
  \\
  \nonumber
   & 
  =
  \frac{1}{\imag \hbar} \AcPropagation \hat \phi_\beta + \frac{1}{2 \imag \hbar}
  \sum_{\alpha \alpha'} 
  (\bar \Gamma_{\alpha' \alpha \beta}^\ast 
  \hat \psi_\alpha \hat \psi_{\alpha'}^\dagger 
  + \bar \Gamma_{\alpha \alpha' \beta}^\ast
  \hat \psi_\alpha^\dagger \hat \psi_{\alpha'}).
\end{align}
As we have already discussed, the commutator 
\mbox{$[\hat \psi_{\alpha'}, \hat \psi_\alpha^\dagger]$} is irrelevant for the
dynamics of FBS and therefore it can be safely ignored:
\begin{align}
   \partial_t \hat \phi_\beta = & 
  \frac{1}{\imag \hbar} \AcPropagation \hat \phi_\beta 
  \\
  \nonumber
  & \quad 
  + 
  \frac{1}{2 \imag \hbar} \sum_{\alpha \alpha'} 
  (
  \bar \Gamma_{\alpha' \alpha \beta}^\ast 
  \hat \psi_{\alpha'}^\dagger \hat \psi_{\alpha}
  + 
  \bar \Gamma_{\alpha \alpha' \beta}^\ast 
  \hat \psi_{\alpha}^\dagger \hat \psi_{\alpha'}
  ).
\end{align}

It is convenient to transform the optical operators to a rotating frame that 
eliminates both the temporal and the spatial carrier.
It should be stressed that we do not apply such a transformation to the 
acoustic envelopes as this would lead to an explicitly time-dependent 
interaction term. 
Furthermore, the (nearly) vanishing acoustic group velocity suggests a 
rescaling to normalize with respect to energies rather than powers:
\begin{align}
  \hat \Psi_\alpha(z, t) 
  = & \hat \psi_{\alpha}(z, t) \sqrt{\hbar \omega_0} 
  \exp(\imag \omega_0 t - \imag k_\alpha z),
  \\
  \hat \Phi_\beta(z, t) = & \hat \phi_{\beta}(z, t) 
  \sqrt{\hbar \Omega_{\beta q_\beta}},
  \\
  \Gamma_{\alpha \alpha' \beta} 
  = & \frac{1}{ \hbar \omega_0 \sqrt{\hbar \Omega_{\beta q_\beta}} }
  \bar \Gamma_{\alpha \alpha' \beta}
  =
  \frac{1}{\sqrt{\Energy_{\alpha} \Energy_{\alpha'} \Energy_{\beta}}} Q_{\alpha \alpha' \beta}
  .
\end{align}
The product operators $\hat \Psi^\dagger \hat \Psi$ and 
$\hat \Phi^\dagger \hat \Phi$ thus express the local optical and acoustic 
energies rather than the local number of photons and phonons, respectively.
%\end{widetext}
\begin{widetext}
\noindent
Using Eqs.~(\ref{eqn:Lalpha_def},\ref{eqn:Lbeta_def}), the resulting quantum 
equations of motion are:
\begin{align}
  \nonumber
  \left[ \partial_t + v_\alpha^{(1)} \partial_z - 
  \imag \frac{v_\alpha^{(2)}}{2} \partial_z^2 + \ldots \right] 
  \hat \Psi_\alpha 
  = & - \frac{\imag \omega_0}{2} \sum_{\alpha' \beta} 
  \mathrm{e}^{\imag(k_{\alpha'} - k_\alpha)z}
  \Big[
    (\Gamma_{\alpha \alpha' \beta} + \Gamma_{\alpha' \alpha \beta} )
    \hat \Phi_\beta 
    \\
    & \quad +
    (\Gamma_{\alpha \alpha' \beta}^\ast + \Gamma_{\alpha' \alpha \beta}^\ast )
    \hat \Phi_\beta^\dagger
  \Big] \hat \Psi_{\alpha'} 
  ,
  \label{eqn:eom_psi_qm}
  \\
  \nonumber
  \left[ (\partial_t + \imag \Omega_{\beta q_\beta}) 
  + v_\beta^{(1)} (\partial_z - \imag q_\beta) - 
  \imag \frac{v_\beta^{(2)}}{2} (\partial_z - \imag q_\beta)^2 + 
  \ldots
  \right]
  \hat \Phi_\beta 
  = & - \frac{\imag \Omega_{\beta q_\beta}}{2} \sum_{\alpha \alpha'} 
  \Big[
    \mathrm{e}^{\imag(k_{\alpha'} - k_{\alpha})z}
    \Gamma_{\alpha' \alpha \beta}^\ast 
    \hat \Psi_\alpha^\dagger \hat \Psi_{\alpha'}
    \\
    & \quad +
    \mathrm{e}^{\imag(k_{\alpha} - k_{\alpha'})z}
    \Gamma_{\alpha \alpha' \beta}^\ast 
    \hat \Psi_{\alpha'}^\dagger \hat \Psi_\alpha)
  \Big] .
  \label{eqn:eom_phi_qm}
\end{align}

In most situations a classical description is sufficient and allows for the 
simple inclusion of loss via damping coefficients $\gamma_\alpha$ and 
$\gamma_\beta$.
To this end, we identify the expectation value of the QM energy density 
operator with the classical energy density of modulated modes:
\begin{align}
  \langle \hat \Psi_\alpha^\dagger \hat \Psi_\alpha \rangle 
  = & \Energy_\alpha |a_\alpha|^2,
  &
  \langle \hat \Phi_\beta^\dagger \hat \Phi_\beta \rangle 
  = & \Energy_\beta |\tilde b_\beta|^2,
\end{align}
where $a_\alpha(z,t)$ and $\tilde b_{\beta}(z,t)$ are classical 
dimensionless envelope functions.
The tilde indicates that the acoustic envelopes are subject to a different
phase convention than the optical envelopes.
The classical equations of motion are therefore obtained from
Eqs.~(\ref{eqn:eom_psi_qm},\ref{eqn:eom_phi_qm}) by identifying:
\begin{align}
  \hat \Psi_\alpha \rightarrow & \sqrt{\Energy_\alpha} a_\alpha,
  &
  \hat \Phi_\beta \rightarrow & \sqrt{\Energy_\beta} \, {\tilde b}_\beta;
\end{align}
\begin{align}
  \nonumber
  \left[ \gamma_\alpha + \partial_t + v_\alpha^{(1)} \partial_z - 
  \imag \frac{v_\alpha^{(2)}}{2} \partial_z^2 + \ldots \right] 
  a_\alpha
  = & - \frac{\imag \omega_{\alpha k_\alpha}}{\Energy_\alpha} 
  \sum_{\alpha' \beta} 
  \mathrm{e}^{\imag(k_{\alpha'} - k_\alpha)z}
  \\
  & \quad \times \Re\left\{
    \big( Q_{\alpha \alpha' \beta} + Q_{\alpha' \alpha \beta} \big) \tilde b_\beta 
  \right\} a_{\alpha'},
  \label{eqn:eom_psi_cl}
  \\
  \nonumber
  \left[ \gamma_\beta + (\partial_t + \imag \Omega_{\beta q_\beta}) 
  + v_\beta^{(1)} (\partial_z - \imag q_\beta) - 
  \imag \frac{v_\beta^{(2)}}{2} (\partial_z - \imag q_\beta)^2 + 
  \ldots
  \right]
  \tilde b_\beta
  = & - \frac{\imag \Omega_{\beta q_\beta} }{2\Energy_\beta}
  \sum_{\alpha \alpha'} \big(
  \mathrm{e}^{\imag(k_{\alpha'} - k_\alpha)z}
  Q_{\alpha' \alpha \beta}^\ast a^\ast_\alpha a_{\alpha'}
  \\
  & \quad + 
  \mathrm{e}^{\imag(k_{\alpha} - k_{\alpha'})z}
  Q_{\alpha \alpha' \beta}^\ast a^\ast_{\alpha'} a_\alpha 
  \big) .
  \label{eqn:eom_phi_cl}
\end{align}
Note that the optical damping coefficient is defined with respect to amplitudes
and not powers.
Thus, it satisfies $\gamma_\alpha=v\alpha_\text{loss}/2$, where 
$\alpha_\text{loss}$ is the conventional power attenuation coefficient.

\end{widetext}

The framework presented thus far is directly applicable intra-mode FBS -- which 
is the main focus of this paper -- and for non-cascaded inter-mode backward SBS.
The treatment of cascaded inter-mode FBS and intra-mode backward SBS requires 
\emph{two} carriers either in an acoustic or in an optical mode in order to 
describe forward and backward propagating acoustic waves (in the case of 
inter-mode FBS) or forward and backward propagating optical waves (in the case 
of intra-mode backward SBS).
operators defined via Fourier transformations over half the reciprocal space.

%%%%%%%%%%%%%%%%%%%%%%%%%%%%%%%%%%%%%%%%%%%%%%%%%%%%%%%%%%%%%%%%%%%%%%%%%%%%%%%%
\subsection{Intra-mode FBS}

Often, only one optical mode $\alpha=\alpha'$ and one acoustic mode $\beta$ 
are relevant.
In this case, the coefficients simplify and phase factors disappear:
\begin{align}
  Q_{\alpha \alpha' \beta} = & Q_{\alpha' \alpha \beta} = Q,
  \\
  \mathrm{e}^{\imag(k_{\alpha'} - k_\alpha)z} = &
  \mathrm{e}^{\imag(k_\alpha - k_{\alpha'})z} = 1,
\end{align}
the mode-index summations in the interaction terms disappear
and we drop the subscripts $\alpha$, $\beta$, $k_\alpha$ and $q_\beta$ 
where appropriate and use $k_0$, $\Omega_0$, and $q_0$ to denote the optical
and acoustic carrier wave parameters.
This leads to simplified quantum equations of motion
\begin{align}
  \left[ \partial_t + v_\alpha^{(1)} \partial_z 
  + \ldots \right] \hat \Psi
  = - \imag \omega_0 
  (\Gamma \hat \Phi + 
  \Gamma^\ast \hat \Phi^\dagger) 
  \hat \Psi,
  \\
  \left[ (\partial_t + \imag \Omega_0) 
  + v_\beta^{(1)} (\partial_z - \imag q_0) 
  + \ldots \right] \hat \Phi
  = - \imag \Omega_0 \Gamma^\ast \hat \Psi^\dagger \hat \Psi,
\end{align}
and their classical counterparts:
\begin{align}
  \left[ \gamma_\alpha + \partial_t + v_\alpha^{(1)} \partial_z 
    %+ \imag \frac{v_\alpha^{(2)}}{2} \partial_z^2 
  + \ldots \right] a
  = - \frac{2 \imag \omega_0}{\Energy_\alpha} \Re\{Q \tilde b\} a,
  \\
  \left[ \gamma_\beta + (\partial_t + \imag \Omega_0)
  + v_\beta^{(1)} (\partial_z - \imag q_0) 
  %+ \imag \frac{v_\beta^{(2)}}{2} (\partial_z - \imag q_0)^2 
  + \ldots \right] \tilde b
  = - \frac{\imag \Omega_0}{\Energy_\beta} Q^\ast |a|^2.
\end{align}
The remainder of this paper is entirely based on the classical equations of 
motion with only one relevant optical and one relevant acoustic mode.

We now demonstrate how the Eqs.~(\ref{eqn:eom_psi_cl},\ref{eqn:eom_phi_cl})
can be reduced to the standard treatment with individual optical amplitudes
for each Stokes order.
First, we assume that the optical amplitude can be represented as a 
superposition of a pump and $N$ Stokes and $M$ anti-Stokes orders:
\begin{align}
  a(z,t) = \sum_{n=-M}^N a_n(z,t) \exp[-\imag n (\Omega_0 - q_0 z t)],
\end{align}
where each individual amplitude $a_n$ is assumed to vary slowly on the time
and length scales of the \emph{acoustic} problem.
Furthermore, it is now convenient to transform the acoustic envelope to a 
rotating frame:
\begin{align}
  b(z,t) = \tilde b(z, t) \exp(\imag \Omega_0 t - \imag q_0 z).
\end{align}
We insert this into Eqs.~(\ref{eqn:eom_psi_cl}) to obtain:
%\begin{widetext}
%\begin{align}
%  \nonumber
%  & \sum_{n=-M}^N 
%  \mathrm{e}^{\imag n (q_0 z - \imag \Omega_0 t)}
%  \left[ \gamma_\alpha + (\partial_t - \imag n \Omega_0) + 
%    v_\alpha^{(1)} (\partial_z + \imag n q_0) - 
%  \imag \frac{v_\alpha^{(2)}}{2} (\partial_z^2 + 2 n \imag q_0 - q_0^2) 
%  + \ldots \right] a_n
%  \\
%  = & \frac{\imag \omega_0}{\Energy_\alpha} \sum_{n=-M}^N \left[
%    \mathrm{e}^{\imag (n + 1) (q_0 z - \imag \Omega_0 t)} 
%    Q b a_n + 
%    \mathrm{e}^{\imag (n - 1) (q_0 z - \imag \Omega_0 t)} 
%    Q^\ast b^\ast a_n
%  \right].
%\end{align}
%\end{widetext}
\begin{align}
  0 = & \sum_{n=-M}^N a_n
  \Bigg\{
  \mathrm{e}^{\imag n (\imag \Omega_0 t - q_0 z)}
  \Bigg[ \gamma_\alpha + (\partial_t - \imag n \Omega_0) 
    \\
    \nonumber 
    & 
    + v_\alpha^{(1)} (\partial_z + \imag n q_0) 
    - \imag \frac{v_\alpha^{(2)}}{2} (\partial_z^2 + 2 n \imag q_0 - q_0^2) 
    + \ldots \Bigg] 
    \\
    \nonumber 
    & 
    + \frac{\imag \omega_0}{\Energy_\alpha} \Bigg[
    \mathrm{e}^{\imag (n + 1) (\imag \Omega_0 t - q_0 z)} 
    Q b 
    + 
    \mathrm{e}^{\imag (n - 1) (\imag \Omega_0 t - q_0 z)} 
    Q^\ast b^\ast 
  \Bigg] \Bigg\}.
\end{align}
Given the slowly varying constraint,
this equation is equivalent to a family of equations for the individual Stokes
orders:
\begin{align}
  \nonumber
  & 
  \left[ \gamma_\alpha + (\partial_t - \imag n \Omega_0) + 
    v_\alpha^{(1)} (\partial_z + \imag n q_0) + 
  %\imag \frac{v_\alpha^{(2)}}{2} (\partial_z^2 + 2 n \imag q_0 - q_0^2) + 
  \ldots \right] a_n
  \\
  = & - \frac{\imag \omega_0}{\Energy_\alpha} (Q b a_{n-1} + Q^\ast b^\ast a_{n+1}),
  \label{eqn:expanded_psi}
\end{align}
with the truncation condition
\begin{align}
  a_{-M-1} = & a_{N+1} = 0.
  \label{eqn:expanded_truncation_condition}
\end{align}
Analogously, we now insert the optical decomposition into the acoustic equation
of motion~\eqref{eqn:eom_psi_cl}:
\begin{align}
  & \left[ \gamma_\beta + \partial_t + v_\beta^{(1)} \partial_z - 
  \imag \frac{v_\beta^{(2)}}{2} \partial_z^2 + \ldots \right]
  b
  \label{eqn:expanded_phi}
  \\
  \nonumber
  = & - \frac{\imag \Omega_0 Q^\ast}{\Energy_\beta} \sum_{n=-M}^{N-1} 
  a_n^\ast a_{n+1} + \text{non-resonant terms},
\end{align}
where ``non-resonant terms'' contains all products of optical envelopes where
the beating is not proportional to $\exp(\imag q_0 z - \imag \Omega_0 t)$.
Neglecting these terms is equivalent to tightening the optical RWA to an
acousto-optical RWA that ignores all contributions that disappear when averaged
over a time scale much longer than the \emph{acoustic} period.
Eqs.~(\ref{eqn:expanded_psi}--\ref{eqn:expanded_phi}) are the conventional 
coupled-mode equations for a large number of (anti-) Stokes orders.

The common case of a pump and a single Stokes side-band are obtained for $M=0$, $N=1$:
\begin{align}
  \left( \gamma_\alpha + \partial_t + v_\alpha^{(1)} \partial_z 
    %+ \imag \frac{v_\alpha^{(2)}}{2} \partial_z^2 
  + \ldots \right) a_0
  = & - \frac{\imag \omega_0 Q}{\Energy_\alpha} b a_1,
  \label{eqn:dual_mode_1}
  \\
  \left( \gamma_\alpha + \partial_t + v_\alpha^{(1)} \partial_z 
    %+ \imag \frac{v_\alpha^{(2)}}{2} \partial_z^2 
  + \ldots \right) a_1
  = & - \frac{\imag \omega_0 Q^\ast}{\Energy_\alpha} b^\ast a_0,
  \label{eqn:dual_mode_2}
  \\
  \left( \gamma_\beta + \partial_t + v_\beta^{(1)} \partial_z 
    %+ \imag \frac{v_\beta^{(2)}}{2} \partial_z^2 
  + \ldots \right) b
  = & - \frac{\imag \Omega_0 Q^\ast}{\Energy_\beta} a_1^\ast a_0,
  \label{eqn:dual_mode_3}
\end{align}
where we have used the fact that the terms containing $\Omega_0$ and $q_0$ 
cancel each other in Eq.~\eqref{eqn:dual_mode_2} if the acoustic wave satisfies
the phase-matching condition.
The terms with $\Omega_0$ and $q_0$ never appear in Eq.~\eqref{eqn:dual_mode_1},
because $n=0$.

Equations~(\ref{eqn:dual_mode_1}--\ref{eqn:dual_mode_3}) are in agreement with
the literature on conventional classical coupled mode theory for 
SBS~\cite{Rakich2012,Wolff2015}.
With this, we have demonstrated that the commonly employed multi-envelope theory
to FBS emerges from our treatment by restricting the assumptions about the 
optical spectrum and neglecting terms oscillating at multiples of the acoustic 
frequency in the interaction.

%%%%%%%%%%%%%%%%%%%%%%%%%%%%%%%%%%%%%%%%%%%%%%%%%%%%%%%%%%%%%%%%%%%%%%%%%%%%%%%%
%%%%%%%%%%%%%%%%%%%%%%%%%%%%%%%%%%%%%%%%%%%%%%%%%%%%%%%%%%%%%%%%%%%%%%%%%%%%%%%%
\section{Complete analytical solution in dispersionless intra-mode FBS}
\label{sec:dispersionless}

Optical and acoustic dispersion are typically very small over the narrow 
frequency ranges (usually MHz to GHz) of Brillouin phenomena. 
Furthermore the propagation length of sound in a waveguide can be far smaller 
than the variation of the acoustic and optical envelopes, and so the 
propagation of acoustic waves is often neglected in Brillouin calculations. 
Having established the governing 
equations~(\ref{eqn:eom_psi_cl},\ref{eqn:eom_phi_cl})
we now examine the important limit where both optical and acoustic dispersion 
can be entirely neglected, and where the propagation length of acoustic waves 
is assumed to be equal to zero. 
We will see that this canonical situation provides insight, in the form of 
analytic solutions to the governing equations, into the overall behavior of 
FBS processes.
In this section, we will solve the dispersionless initial value problem
corresponding to Eqs.~(\ref{eqn:eom_psi_cl},\ref{eqn:eom_phi_cl}) in this 
no-dispersion, no-acoustic-propagation limit:
\begin{align}
  \left( \partial_t + v \partial_z \right) a
  = & - \left(\frac{2 \imag \omega_0}{\Energy_\alpha} 
  \Re\{Q \tilde b\} + \gamma_\alpha 
  \right) a,
  \label{eqn:anasol_eom_opt}
  \\
  \left( \partial_t + \gamma_\beta + \imag \Omega_0 \right) \tilde b
  = & - \frac{\imag \Omega_0 Q^\ast}{\Energy_\beta} |a|^2,
  \label{eqn:anasol_eom_ac}
  \\
  a(z=0, t) = & A(t),
  \label{eqn:anasol_eom_bc}
  \\
  a(z, t) = & \tilde b(z, t) = A(t) = 0 \quad \text{for} \quad t < 0, 
  \label{eqn:anasol_eom_ic}
\end{align}
where $A(t)$ is the incident optical envelope at the start of the waveguide.

The general approach in solving this system is to first integrate the optical
equation~\eqref{eqn:anasol_eom_opt} to find the optical amplitude as a function
of $z$ and $t$. 
This solution will depend on the incident optical field $A(t)$ and the yet 
unknown acoustic envelope $\tilde b(z, t)$.
We then insert it into Eq.~\eqref{eqn:anasol_eom_ac} and find that the 
resulting equation happens to be solvable.
By back-substitution into the solution to Eq.~\eqref{eqn:anasol_eom_opt} we
find the analytical solution to cascaded FBS as an explicit integral and 
evaluate it for a typical experimental setup.

%%%%%%%%%%%%%%%%%%%%%%%%%%%%%%%%%%%%%%%%%%%%%%%%%%%%%%%%%%%%%%%%%%%%%%%%%%%%%%%%
\subsection{General solution}

Eq.~\eqref{eqn:anasol_eom_opt} is separable when integrating along the 
characteristic of the one-way wave equation.
In other words, we obtain the phase of the optical amplitude at the point
\mbox{$(z,t)$} by integrating along a line \mbox{$(z',t')$} in space time 
defined by
\begin{align}
  z' = & z - vs, & t' = & t - s,
\end{align}
where $s$ is the running parameter.
%\begin{align}
%  y = & v t + z; \quad \Rightarrow \quad \partial_y = 
%  \frac{1}{v} \partial_t + \partial_z,
%  \\
%  \frac{\partial_y a}{a} = & - \frac{2 \imag \omega_0}{v \Energy_\alpha}
%  \Re\{Q \tilde b(y, t-y/v)\} - \frac{\gamma_\alpha}{v},
%\end{align}
This leads to the solution (expressed in the original coordinates):
\begin{align}
  a(z, t) = & E\Big(t - \frac{z}{v}\Big) 
  \exp\Big(-\frac{\gamma_\alpha z}{ v}\Big)
  \label{eqn:anasol_step1}
  \\
  \nonumber & \ \times
  \exp\left[\frac{2 \omega_0}{\imag \Power_\alpha}\int_0^z \total z' \ 
    \Re\left\{Q \tilde b\left(z', t - \frac{z - z'}{v}\right)\right\}
  \right].
\end{align}
It should be pointed out that the integrand is purely real-valued and thus
the acoustic envelope only introduces a phase.
In the absence of optical dispersion, the effect of the acoustic deformation on 
the optical wave is a pure phase modulation and the optical intensity 
does not depend on $\tilde b$.
However, the only thing relevant for Eq.~\eqref{eqn:anasol_eom_ac} is the 
optical intensity distribution:
\begin{align}
  |a(z, t)|^2 = & \left|E\left(t - \frac{z}{v}\right)\right|^2 
  \exp\Big(-\frac{2 \gamma_\alpha z }{ v} \Big).
\end{align}
With this, Eq.~\eqref{eqn:anasol_eom_ac} becomes an ordinary differential 
equation that can be explicitly solved via its Green function:
\begin{align}
  \tilde b(z, t) = & \frac{\Omega_0 Q^\ast}{\imag \Energy_\beta} 
  \int_0^\infty \total t' \ 
  \text{e}^{-(\gamma_\beta + \imag \Omega_0)t' - 
  \frac{2\gamma_\alpha z}{v}}
  \left| E\left(t - t' - \frac{z}{v}\right) \right|^2,
  \label{eqn:anasol_step2}
\end{align}
This expression only depends on $A(t)$, but not on $a(z, t)$.
Therefore, Eq.~\eqref{eqn:anasol_step1} is a simple integral rather than an
integral equation.
Furthermore, the $z$-integral in Eq.~\eqref{eqn:anasol_step1} is trivial.
We find for the $\tilde b$-related term appearing under the $z$-integral:
\begin{align}
  & - \Re\left\{ Q \tilde b\left(z', t - \frac{z - z'}{v}\right)\right\}
  \\
  \nonumber
  = & \frac{\Omega_0 |Q|^2}{\Energy_\beta} \int_0^\infty \total t' \ 
  \text{e}^{-\gamma_\beta t' - \frac{2\gamma_\alpha z'}{v}}
  \left| E\left(t - t' - \frac{z}{v}\right) \right|^2 
  \sin(\Omega_0 t').
\end{align}
The stationary solution for the optical envelope of cascaded FBS in the 
dispersionless case and for any incident optical envelope $A(t)$ is therefore:
\begin{align}
  \nonumber
  & a(z, t) = E\bigg(t - \frac{z}{v}\bigg) 
  \exp\bigg\{-\frac{\gamma_\alpha z}{v} 
    + \frac{\imag \Omega_0 \omega_0 |Q|^2 
    \big(1 - \text{e}^{-\frac{2 \gamma_\alpha z}{v}}\big)}{
      \gamma_\alpha \Energy_\alpha \Energy_\beta
    }
    \\
    \nonumber
    & \quad \times 
    \int_0^\infty \total t' \ \text{e}^{-\gamma_\beta t'}
    \left| E\left(t - t' - \frac{z}{v}\right) \right|^2 
    \sin(\Omega_0 t')
  \bigg\},
  \\
  \nonumber
  &  = E\bigg(t - \frac{z}{v}\bigg)  
  \exp\bigg\{-\frac{\gamma_\alpha z}{v} 
    + \frac{\imag  \Omega_0 \omega_0 |Q|^2 }{ \Energy_\alpha \Energy_\beta }
    \frac{2 L_\text{eff}}{v}
    \\
    & \quad \times 
    \int_0^\infty \total t' \ \text{e}^{-\gamma_\beta t'}
    \left| E\left(t - t' - \frac{z}{v}\right) \right|^2 
    \sin(\Omega_0 t')
  \bigg\},  
  \label{eqn:dispersionless_solution}
\end{align}
where the effective propagation length of the optical field at point $z$ is
\begin{align}
  L_\text{eff}=\frac{
  v \big(1 - \text{e}^{-\frac{2 \gamma_\alpha z}{v}}\big)}{ 2 \gamma_\alpha }.
\end{align}
The resulting optical field can then be understood as the input field $A(t)$ 
attenuated according to the damping $\gamma_a/v$ while being phase modulated by 
its own intensity over the effective length.

%%%%%%%%%%%%%%%%%%%%%%%%%%%%%%%%%%%%%%%%%%%%%%%%%%%%%%%%%%%%%%%%%%%%%%%%%%%%%%%%
\subsection{Example: two-tone excitation and zero optical loss}
\label{sec:two-tone}

The explicit formula~\eqref{eqn:dispersionless_solution} is the complete 
solution to the problem of cascaded FBS in the absence of optical dispersion.
To gain more insight into its nature,
we next evaluate it for an example that is typical both for theoretical and
experimental studies of SBS: the excitation with a strong optical pump and a
weak Stokes side band:
\begin{align}
  A(t) = A_0 [1 - \imag M \exp(\imag \Omega_0 t)],
\end{align}
where $A_0$ is the amplitude of the pump wave and $M A_0$ is that of the
Stokes seed.
The corresponding pump and Stokes seed powers are
\begin{align}
  P_\text{P} = & \Power_\alpha |A_0|^2
  &
  P_\text{S} = & \Power_\alpha |M A_0|^2 
\end{align}
We assume that $M$ is real-valued, which is always possible by an
appropriately chosen time origin.
We furthermore neglect the optical loss $\gamma_\alpha$ in this specific 
example; the inclusion is straight-forward but tends to obscure the 
equations.
On thus letting $L_\text{eff}\to z$ and $\gamma_\alpha \to 0$,
the solution becomes
%\begin{widetext}
\begin{align}
  a(z, t) = & 
  A\Big(t - \frac{z}{v}\Big) 
  \exp\Bigg\{\frac{2 \imag \Omega_0 \omega_0 |Q|^2 z}{\Power_\alpha \Energy_\beta}
    \\
    \nonumber
    & \quad \times 
    \int_0^\infty \total t' \ \text{e}^{-\gamma_\beta t'}
    \left| A\Big(t - t' - \frac{z}{v}\Big) \right|^2 
    \sin(\Omega_0 t')
  \Bigg\}
  \\
%  = &
%  f\big(t - \frac{z}{v}\big) 
%  \exp\left\{\frac{2 \imag \Omega_0 \omega_0 |Q|^2 |f_0|^2 z}{\Power_\alpha \Energy_\beta}
%    \int_0^\infty \total t' \ \text{e}^{-\gamma_\beta t'}
%    \left(
%      1 + |M|^2 + 2 M \sin\Big[
%    \Omega_0 \Big(t - t' - \frac{z}{v}\Big) \Big]
%    \right)
%    \sin(\Omega_0 t')
%  \right\}
%  \\
  = &
  %A_0 \bigg( 1 + M \mathrm{e}^{\imag \Omega_0 (t - \frac{z}{v})} \bigg)
  A\Big(t - \frac{z}{v}\Big) 
  \exp\Bigg\{
    \underbrace{
      \frac{2 \imag \Omega_0^2 \omega_0 |Q|^2 (P_\text{P} + P_\text{S}) z}{
      \Power_\alpha^2 \Energy_\beta (\gamma_\beta^2 + \Omega_0^2) } 
    }_{\text{static term}} 
      \\ & \quad +
    - \frac{2 \imag \Omega_0 \omega_0 |Q|^2 \sqrt{P_\text{P} P_\text{S}} z}{\gamma_\beta \Power_\alpha^2 \Energy_\beta} 
    \Bigg[
    \nonumber
    \underbrace{
      \cos\Big(\Omega_0 t - \frac{\Omega_0 z}{v} \Big)
    }_{\text{dominant term}}
    \\
    \nonumber
    & \quad +
    \underbrace{ 
      \frac{\gamma_\beta \mathrm{e}^{-\imag \Omega_0 (t - z/v)} }{2 (\gamma_\beta - 2 \imag \Omega_0)}
      + \frac{\gamma_\beta \mathrm{e}^{\imag \Omega_0 (t - z/v)} }{2 (\gamma_\beta + 2 \imag \Omega_0)}
    }_{\text{counter-rotating term}}
  \Bigg] \Bigg\}.
\end{align}
%\end{widetext}
For high mechanical quality factors (\ie $\Omega_0 \gg \gamma_\beta$), the 
additional phase factor due to static waveguide deformations and the 
out-of-phase contributions due to the counter-rotating term are negligible 
compared to the resonant effect.
We introduce a natural unit of length that depends on the optical input powers
and the intrinsic SBS parameters and we shift the origin of the time coordinate 
to a more convenient point:
\begin{align}
  \zeta = & z / L_\text{nat},
  \\
  \tau = & t - \frac{z}{\Omega_0 v},
  \\
  \text{with} \quad L_\text{nat} = & 
  \frac{\Power_\alpha^2 \Energy_\beta \gamma_\beta}{2 \Omega_0 \omega_0 |Q|^2 \sqrt{P_\text{P} P_\text{S}} }.
  \label{eqn:units}
\end{align}
In these units, the cascaded FBS effect [excluding static deformations that
only introduce a very slow phase factor $\exp(\imag A \zeta)$ where $A$ is
approximately the inverse mechanical quality factor]
takes a generic and simple form:
\begin{align}
  a(\zeta, \tau) = & A(\tau) \exp[\imag \zeta \cos(\Omega_0 \tau)].
  \label{eqn:dispersionless_solution_natunits}
\end{align}

We will now investigate three properties of this solution that are of particular
interest for the characterization of SBS-based frequency combs: the frequency 
spectrum, the intensity evolution and the auto-correlation function.

\begin{figure}
  \includegraphics[width=\columnwidth]{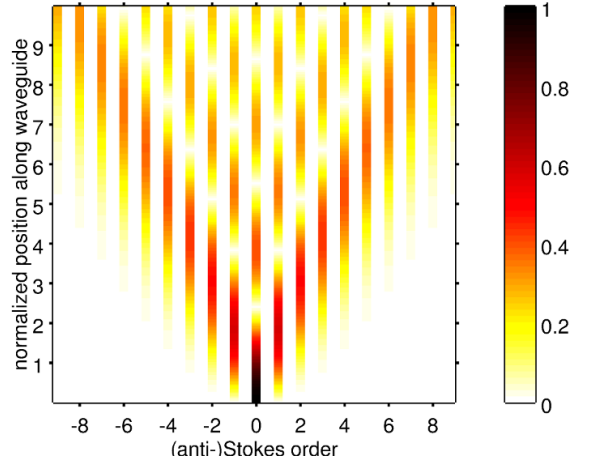}
  \caption{
    Evolution of the absolute amplitude (gray scale) of various \mbox{(anti-)}Stokes orders 
    (frequency on the abscissa) as a function of position $\zeta$ along the 
    wave guide (ordinate) assuming vanishing dispersion and an optical 
    excitation with a strong pump and a weak Stokes side-band.
  }
  \label{fig:stokes_evolution}
\end{figure}

The frequency spectrum of the total 
signal Eq.~\eqref{eqn:dispersionless_solution_natunits} is the convolution of 
the spectrum of the incident signal $A(\tau)$ and the SBS-modulation function
\begin{align}
  \mathcal{R}(\zeta, \tau) = \exp[\imag \zeta \cos(\Omega_0 \tau)],
\end{align}
which reflects the effect of cascaded FBS.
Due to the periodicity of the problem, its spectrum only contains narrow lines
at integer multiples of the Stokes shift $\Omega_0$.
Therefore, the spectral component of the $n$-th (anti-)Stokes order is given by
\begin{align}
  \mathcal R_n(\zeta) = & \frac{1}{2\pi} \int_0^{2 \pi / \Omega_0} \total \tau \ 
  \exp[\imag n \Omega_0 \tau + \imag \zeta \cos(\Omega_0 \tau)]
  \\
  = & \frac{1}{\Omega_0} J_n(\zeta),
\end{align}
where $J_n(\zeta)$ is the Bessel function of order $n$.
We therefore can directly plot the generation of the individual Stokes orders 
along the waveguide (see Fig.~\ref{fig:stokes_evolution}).
Cascaded FBS clearly generates a frequency comb in the sense that it generates
many equally spaced frequency lines.

However, very often frequency combs are associated with the generation of 
light pulses that are equally spaced in time domain.
The solution of cascaded FBS in the absence of optical dispersion in a 
straight waveguide does not predict the formation of pulses.
Even more, the SBS response function $\mathcal{R}(\zeta, \tau)$ is a pure
phase factor, \ie the FBS process does not affect the intensity distribution
at all.
A related feature of this is that (in the absence of optical loss and 
dispersion) the acoustic amplitude is constant along the waveguide: the sound 
wave is not amplified.
This is in agreement with an earlier work on the optical single-pass gain of
FBS structures~\cite{Rakich2016}.
As a result, the effect of cascaded FBS in the non-dispersive regime cannot 
be observed in the RF-domain by detecting the interference within the optical
output with a photodetector.

\begin{figure}
  \includegraphics[width=\columnwidth]{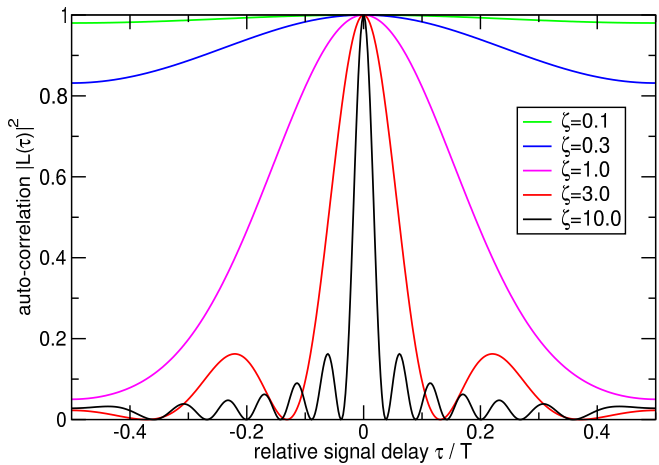}
  \caption{
    Absolute square $|\mathcal{L}(\tau)|^2$ of the auto-correlation function
    [see Eq.~\eqref{eqn:auto_correlation_def}]
    of the SBS-generated signal modulation $\mathcal{R}(\zeta, \tau)$ for 
    a range of waveguide lengths $\zeta$ expressed in natural units of length
    $L_\text{nat}$ [see Eq.~\eqref{eqn:units}].
    The delay parameter $\tau$ is normalized to the acoustic period 
    $T = 2\pi/\Omega_0$.
  }
  \label{fig:auto_correlation}
\end{figure}

Finally, auto-correlation measurements of optical combs are often used instead
of measuring the time-dependence of the optical intensity in order to grade the
quality of a frequency comb.
It turns out that the cascaded FBS process in a dispersionless straight 
waveguide causes a pronounced auto-correlation peak despite the complete lack 
of intensity modulation.
The amplitude auto-correlation for this problem with time period 
$T=2\pi/\Omega_0$ can be defined as:
\begin{align}
  \mathcal{L}(\tau) = & \frac{1}{T} \int_0^T \total \tau' \ 
  \mathcal{R}^\ast(\zeta, \tau') \mathcal{R}(\zeta, \tau' + \tau)
  \label{eqn:auto_correlation_def}
  \\
  \nonumber
  = & \frac{1}{T} \int_0^T \total \tau' \ 
  \exp\big\{
    \imag \zeta [\cos \Omega_0 \tau'  - \cos\Omega_0(\tau'+\tau)]
  \big\}
  %\\
  %\nonumber
  %= & \frac{1}{T} \int_0^T \total \tau' \ 
  %\exp\big\{
  %  \imag \zeta [\cos \Omega_0 (\tau' - \tau/2) 
  %    \\
  %  & \quad - \cos\Omega_0(\tau'+\tau/2)]
  %\big\}
  %\\
  %= & \frac{1}{T} \int_0^T \total \tau' \ 
  %\exp\big[
  %\imag \zeta \sin (\Omega_0 \tau') \sin (\Omega_0 \tau/2)
  %\big]
  \\
  = & J_0\left(2 \zeta \sin\frac{\Omega_0 \tau}{2}\right),
  \label{eqn:auto_correlation}
\end{align}
where $J_0(x)$ is the Bessel function of order zero.
In Fig.~\ref{fig:auto_correlation} we show the absolute value squared of this
result for a range of normalized waveguide lengths $\zeta$.

%%%%%%%%%%%%%%%%%%%%%%%%%%%%%%%%%%%%%%%%%%%%%%%%%%%%%%%%%%%%%%%%%%%%%%%%%%%%%%%%
%%%%%%%%%%%%%%%%%%%%%%%%%%%%%%%%%%%%%%%%%%%%%%%%%%%%%%%%%%%%%%%%%%%%%%%%%%%%%%%%
\section{Intra-mode FBS in dispersive waveguides}
\label{sec:dispersive}

In the previous section, we have shown that in the absence of optical dispersion
our cascaded model of FBS does not exhibit variations in the acoustic amplitude 
nor the intensity modulation along the waveguide.
This is in fact not due to some hidden assumptions within our model but the
result of the conservation of momentum and energy, \ie phase matching.
To illustrate this, we assume a waveguide that is acoustically modulated 
with the angular frequency $\Omega_0$ and ask how the forward-scattering of 
light by this modulation modifies the acoustic intensity.

\begin{figure}
  \includegraphics[width=0.7\columnwidth]{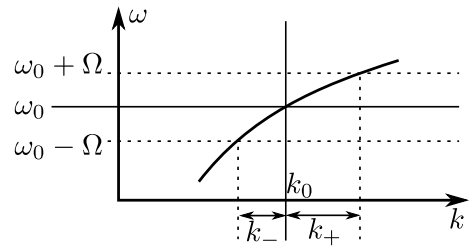}
  \caption{
    Sketch of a dispersive optical band.
    Three points are highlighted: a carrier wave at $\omega_0$ and $k_0$,
    an anti-Stokes wave at $\omega_0 + \Omega_0$ and $k_0 + k_+$ and a
    Stokes wave at $\omega_0 - \Omega_0$ and $k_0 + k_-$ (note that $k_-$ is
    chosen to be negative).
  }
  \label{fig:dispersion_sketch}
\end{figure}

Fig.~\ref{fig:dispersion_sketch} shows schematically a small section of an 
optical dispersion relation: a pump field at angular frequency $\omega_0$ and 
wave number $k_0$ is scattered into the Stokes and anti-Stokes side bands at 
angular frequencies $\omega_0 \mp \Omega_0$ and wave numbers $k_0 + k_\mp$.
The wave number differences can be expanded in terms of the frequency:
\begin{align}
  k_\pm = & \pm s^{(1)} \Omega_0 + \frac{s^{(2)}}{2} \Omega_0^2 + \ldots ,
\end{align}
where the group slowness and slowness dispersion parameters are connected to 
the more familiar group velocity and dispersion parameters 
[Eq.~\eqref{eqn:group_velocity_def}] via: 
\begin{align}
  s^{(1)} = & [v^{(1)}]^{-1}, & s^{(2)} = -v^{(2)}[v^{(1)}]^{-3}.
\end{align}
When part of the optical pump at $\omega_0$ is scattered, its energy and
momentum are distributed between the acoustic system and the two optical side 
bands expressed by amplitudes $a_-$ and $a_+$ at the Stokes and anti-Stokes 
frequencies, respectively.

First, we will find a condition for case that the acoustic amplitude is not 
modified, \ie that the energy and momentum are distributed only among the
optical side bands:
\begin{align}
  \hbar k_+ |a_+|^2 + \hbar k_- |a_-|^2 = & 0,
  \\
  \hbar \Omega_0 |a_+|^2 - \hbar \Omega_0 |a_-|^2 = & 0.
\end{align}
This is only possible if $k_+ = -k_-$, \ie for vanishing optical dispersion
(note that the acoustic wave number $q=k_+$ is not zero).
In addition, we find that $|a_+|^2 = |a_-|^2$, \ie that both side bands are 
equally excited in agreement with Sec.~\ref{sec:dispersionless}.
Finally, we may conclude that in the presence of optical dispersion, a two-wave 
process is not sufficient to simultaneously conserve energy and momentum, \ie 
the acoustic system is required as a third participating wave.

Next, we we introduce $B$ as the (positive or negative) change in the 
acoustic intensity $|b|^2$. 
The conservation of energy requires:
\begin{align}
  \hbar \Omega_0 |a_+|^2 - \hbar \Omega_0 |a_-|^2 + \hbar \Omega_0 B = & 0,
  \\
  \Rightarrow \quad B = & |a_-|^2 - |a_+|^2.
  \label{eqn:qualitative_disp_energy_cons}
\end{align}
Self-amplification occurs if energy is dumped into the acoustic field, \ie
for $B>0$.
For the conservation of momentum, we find:
\begin{align}
  \hbar k_+ |a_+|^2 + \hbar k_- |a_-|^2 + \hbar q B = & 0,
  \\
  \Rightarrow \quad 
  \frac{s^{(2)}\Omega_0^2}{2} (|a_-|^2 + |a_+|^2)  + (q - s^{(1)} \Omega_0) B = & 0.
  \label{eqn:qualitative_disp_momentum_cons}
\end{align}
From this we can see that a change in the acoustic intensity is necessary in 
the presence of optical dispersion.
Furthermore, we can see that (unlike in the dispersionless case) the acoustic 
phase velocity must slightly differ form the optical group velocity.

This very simple argument based on the conservation of energy and momentum 
alone demonstrates that dispersion is necessarily required to modify the 
acoustic intensity via FBS.
However, it cannot predict in which regime self-amplification occurs, because
no assumptions are made about the opto-acoustic interaction.
This requires a more detailed approach, as we will show in the following.

%%%%%%%%%%%%%%%%%%%%%%%%%%%%%%%%%%%%%%%%%%%%%%%%%%%%%%%%%%%%%%%%%%%%%%%%%%%%%%%%
\subsection{Amplification \& suppression in the weak dispersion limit}

We now study the full problems of FBS with optical dispersion, 
where we neglect optical loss for the sake of simplicity:
\begin{align}
  \bigg( \partial_t + v^{(1)}_\alpha \partial_z - 
  \frac{\imag v^{(2)}_\alpha}{2} \partial_z^2 \bigg) a
  = & \frac{2 \imag \omega_0}{\Energy_\alpha} 
  \Re\{Q \tilde b\} a,
  \label{eqn:dispersive_eom_opt_orig}
  \\
  \left( \partial_t + \gamma_\beta + \imag \Omega_0 \right) \tilde b
  = & - \frac{\imag \Omega_0 Q^\ast}{\Energy_\beta} |a|^2,
  \label{eqn:dispersive_eom_ac_orig}
  \\
  a(z=0, t) = & A(t),
  \label{eqn:dispersive_eom_bc_orig}
  \\
  a(z, t) = b(z, t) = A(t) = & 0 \quad \text{for} \quad t < 0.
  \label{eqn:dispersive_eom_ic_orig}
\end{align}
In this problem, the optical dispersion gradually shifts the phase between 
adjacent Stokes lines with propagation along $z$ and therefore slowly 
converts the pure phase-modulation 
of the opto-acoustic interaction into a partial amplitude-modulation.
Depending on the sign of this process, the optical interference pattern is
amplified or attenuated along the waveguide.
Unfortunately, the parabolic nature of the optical equation precludes 
a closed form solution similar to Eq.~\eqref{eqn:dispersionless_solution}.
We therefore restrict ourselves to a few relevant aspects of the problem.

First, we eliminate constants that only clutter the notation by substituting
\begin{align}
  f(z, t) = & \frac{|Q|}{\sqrt{\Energy_\alpha \Energy_\beta}}
  a(z, t),
  \\
  g(z, t) = & - \frac{Q}{\Energy_\alpha} \tilde b(z, t),
\end{align}
and by introducing the symbols $v=v_\alpha^{(1)}$ and $w = v_\alpha^{(2)}/2$.
This leads to the equivalent equations:
\begin{align}
  [\imag \partial_t + \imag v \partial_z + w \partial_z^2 + \omega_0 (g + g^\ast)] f = & 0,
  \label{eqn:dispersive_eom_opt_rescaled}
  \\
  (\partial_t + \gamma_\beta + \imag \Omega_0) g = & \imag \Omega_0 |f|^2.
  \label{eqn:dispersive_eom_ac_rescaled}
\end{align}

In the case of weak dispersion ($w \approx 0$), an approximate solution to
Eqs.~(\ref{eqn:dispersive_eom_opt_rescaled},\ref{eqn:dispersive_eom_ac_rescaled})
over a short length of waveguide ($z\approx0$) can be found via first order
perturbation theory.
To this end, we first make the ansatz
\begin{align}
  f(z, t) = \exp[h(z, t) + \imag j(z, t)],
\end{align}
with real-valued functions $h(z, t)$ and $j(z, t)$.
Inserting this into Eq.~\eqref{eqn:dispersive_eom_opt_rescaled} and 
separating real and imaginary parts of this equation, we reach the system
\begin{align}
  (\partial_t + \gamma_\beta + \imag \Omega_0) g = & \imag \Omega_0 \text{e}^{2 h},
  \\
  \partial_t h + [v + 2 w (\partial_z j)] \partial_z h = & - w \partial^2_z j,
  \\
  (\partial_t + v \partial_z) j + w [ (\partial_z j)^2 - (\partial_z h)^2 
  - \partial_z^2 h] = & \omega_0 (g + g^\ast).
\end{align}
We then write the exact function $h$ ($j$ and $g$ likewise) as an expansion:
\begin{align}
  h(z, t) = h_0(z, t) + h_1(z, t) + h_2(z, t) + \ldots\ ,
\end{align}
where $h_0$ is the solution to the dispersionless problem, $h_n$ are the 
$n^\text{th}$-order corrections corresponding to the powers $w^n$, and we 
will truncate after the first order term.
As in the dispersionless solution, we take our previous result from 
Sec.~\ref{sec:two-tone}:
\begin{align}
  g_0(z, t) = & \imag G_0 \exp\Big(- \imag \Omega_0 \frac{vt - z}{v} \Big),
  \\
  h_0(z, t) = & \log|F_0| + \log\Big|1 - \imag M
  \text{e}^{\frac{\imag \Omega_0 (v t - z)}{v}} \Big|,
  \\
  j_0(z, t) = & \frac{2 G_0 \omega_0 z}{v} \sin\Big(\Omega_0 \frac{vt - z}{v} \Big),
\end{align}
where 
\begin{align}
  F_0 = \frac{|Q| A_0 }{\sqrt{\Energy_\alpha \Energy_\beta}} 
\end{align}
is the transformed optical mean amplitude and 
we have omitted overall meaningless phase factors due to static deformation
in order to keep the equations concise.
The intensity modulation depth $M$ is assumed to be small and connected to
the acoustic amplitude $G_0 = \imag|F_0|^2 M \Omega_0 / \gamma_\beta$.

Next, we calculate the first-order correction to $h$ and its effect on $g$;
the resulting correction to $j$ is not relevant within this context and 
therefore not shown.
The correction to $h$ satisfies:
\begin{align}
  \partial_t h_1 + [v + 2 w (\partial_z j_0)] \partial_z h_1 = & - w \partial_z^2 j_0,
\end{align}
with the boundary condition at $z=0$
\begin{align}
  h_1(0, t) = 0.
\end{align}
\begin{figure}
  \includegraphics[width=\columnwidth]{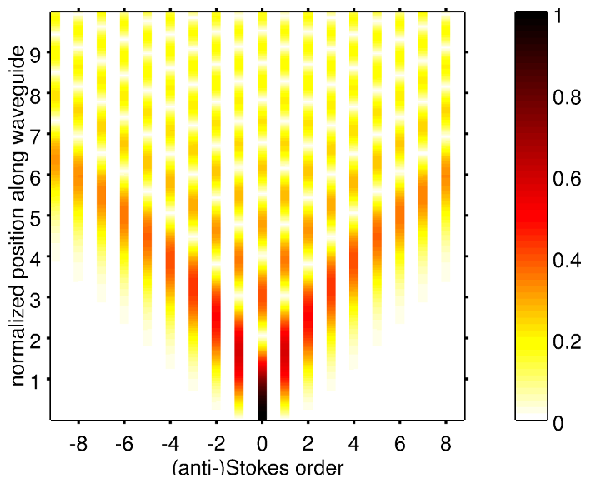}
  \caption{
    Evolution of the \mbox{(anti-)}Stokes orders for a normalized dispersion 
    of $\bar w = -0.15$ assuming that the waveguide is long compared to the 
    acoustic wave length and an optical excitation with a strong pump and a 
    weak Stokes side-band (see also Fig.~\ref{fig:stokes_evolution}).
  }
  \label{fig:neg_dispersion_evolution}
\end{figure}
This leads to the first order correction expressions (the intermediate steps 
leading up to this result can be found in Appendix~\ref{appx:perturbation}):
\begin{align} \label{eq:h1corr}
  h_1(z, t) = & \frac{w \omega_0 \Omega_0 G_0}{v^4} \Big[
  4 vz \cos\Big(\Omega_0 \frac{vt-z}{v}\Big) 
    \\
    \nonumber \quad &
    + \Omega_0 z^2 \sin\Big(\Omega_0 \frac{vt-z}{v}\Big) 
  \Big],
  \\ \label{eq:g1corr}
  g_1(z, t) \approx &
  \frac{2 \imag w \omega_0 \Omega_0^2 G_0 |F_0|^2}{v^4 \gamma_\beta} 
  \Big[ 4 v z + \imag \Omega_0 z^2 \Big] \text{e}^{\imag \Omega_0 \frac{z - vt}{v}},
\end{align}
where we have neglected any non-resonant excitation of the mechanical system.
This is justified for high mechanical quality factors.

This means that to first approximation and over a small distance $z$ the 
acoustic amplitude is modified by a relative factor of
\begin{align}
  \frac{g_1(z, t)}{g_0(z, t)} = 
  \frac{2 w \omega_0 \Omega_0^2 |F_0|^2 ( 4 v z + \imag \Omega_0 z^2 )}{v^4 \gamma_\beta} .
\end{align}
This suggests to make the ansatz
\begin{align}
  g(z, t) = & G(z) g_0(z, t),
  \\
  \text{with} \quad G(\Delta z) \approx &
  \frac{2 w \omega_0 \Omega_0^2 |F_0|^2 ( 4 v \Delta z + \imag \Omega_0 \Delta z^2 )}{v^4 \gamma_\beta}
  \\
  = & (a \Delta z + b \Delta z^2).
\end{align}
for infinitesimally small $\Delta z$.
We can then approximate a waveguide of finite length $z$ as a concatenation of 
pieces of decreasing length $\Delta z$:
\begin{align}
  G(n \Delta z) \approx & (1 + a \Delta z + b \Delta z^2) G[(n-1) \Delta z],
  \\
  \Rightarrow G(z) = & \lim_{z \rightarrow 0} 
  [1 + a\Delta z + b \Delta z^2]^{z/\Delta z}
  \\
  %= & \lim_{z\rightarrow 0} \{\exp[a \Delta z + b\Delta z^2 + \mathcal{O}(\Delta z^2)]\}^{z/\Delta z}
  %\\
  = & \exp\Big( \frac{8 w \omega_0 \Omega_0^2 |F_0|^2}{v^3 \gamma_\beta} z \Big).
  \\
  = & \exp\Big( 
  \frac{8 w \omega_0 \Omega_0^2 |Q|^2 |A_0|^2}{\Power_\alpha \Energy_\beta v^2 \gamma_\beta} z \Big).
  \label{eqn:dispersive_correction}
\end{align}
\begin{figure}
  \includegraphics[width=\columnwidth]{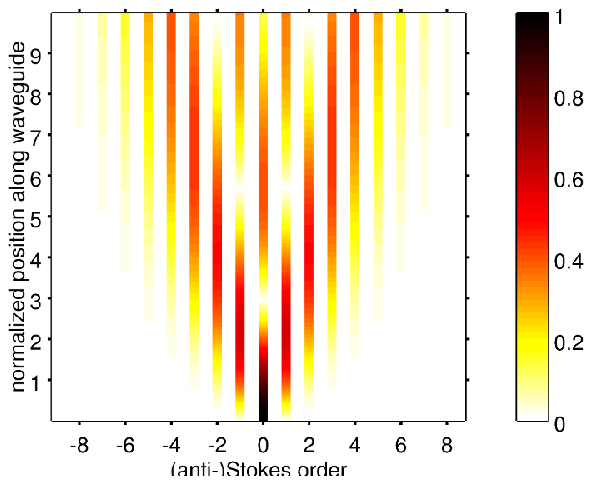}
  \caption{
    Evolution of the \mbox{(anti-)}Stokes orders for a normalised dispersion 
    of $\bar w = 0.15$ assuming that the waveguide is long compared to the 
    acoustic wave length and an optical excitation with a strong pump and a 
    weak Stokes side-band (see also Fig.~\ref{fig:stokes_evolution}).
  }
  \label{fig:pos_dispersion_evolution}
\end{figure}
This can be expressed in terms of the normalised length variable $\zeta$ by 
introducing a normalised dispersion parameter $\bar w$:
\begin{align}
  G(\zeta) = & \exp(\bar w \zeta),
  \\
  \text{with} \quad \bar w = & \frac{4 \Omega_0 w}{v^2} 
  \sqrt{\frac{P_\text{P}}{P_\text{S}}} \ .
\end{align}
Fig.~\ref{fig:neg_dispersion_evolution} and 
Fig.~\ref{fig:pos_dispersion_evolution}, show the evolution of the optical 
spectrum along a waveguide for $\bar w = -0.15$ and $\bar w = +0.15$, 
respectively.
This nicely illustrates that -- within the bounds of this first-order 
perturbation theory, \ie weak dispersion, weak optical intensity contrast --
the optical modulation pattern is compressed along $z$ for $w>0$ and 
stagnates for $w<0$.
The underlying reason for this is that anomalous dispersion translates the
opto-acoustic phase modulation into an amplitude modulation with the same
sign as the optical fluctuation that drives the acoustic oscillation.
As a result, both the intensity modulation and the acoustic amplitude grow
along the waveguide.
In contrast, normal dispersion leads to an amplitude modulation that 
diminishes optical intensity fluctuations along the waveguide.

%%%%%%%%%%%%%%%%%%%%%%%%%%%%%%%%%%%%%%%%%%%%%%%%%%%%%%%%%%%%%%%%%%%%%%%%%%%%%%%%
\subsection{
  Connection of intra-mode FBS to Raman scattering and soliton dynamics
}
\label{sec:raman-like}

The formalism presented so far is particularly well suited for the combination
of FBS with other optical nonlinearities, especially Raman-scattering and the
Kerr effect.
This is because the multi-Stokes optical envelope readily covers nonlinear
interference terms between various Stokes orders, which have to be explicitly
included in more conventional coupled mode theory~\cite{Wolff2015c} and would 
clutter the equations for cascaded Brillouin scattering~\cite{Winful2016}.
The dynamics of solitons in optical fibres can be expressed by a retarded
nonlinear Schroedinger equation,~\cite{Gordon1986} where the instantaneous Kerr 
effect and Raman scattering enter through the convolution of the optical 
intensity with a (time-dependent) response function $r(t)$ as the convolution
kernel.
Our FBS equations of motion including this response function and for the 
complete set of mechanical modes read:
\begin{widetext}
\begin{align}
  \left[ \gamma_\alpha + \partial_t + v_\alpha^{(1)} \partial_z 
  + \imag \frac{v_\alpha^{(2)}}{2} \partial_z^2 \right] a(z, t)
  = & 
  \Bigg[ \underbrace{\frac{2 \omega_0}{\imag \Energy_\alpha} 
    \sum_\beta \Re\{Q_\beta \tilde b_\beta(z, t)\}}_{\text{FBS}}
    + \underbrace{\imag \int_0^\infty \total t' \ r(t') |a(z, t-t')|^2}_{\text{Raman}} \Bigg] 
    a(z, t),
  \\
  \left[ \gamma_\beta + (\partial_t + \imag \Omega_\beta)
  + v_\beta^{(1)} (\partial_z - \imag q_0) \right] \tilde b_\beta(z, t)
  = & - \frac{\imag \Omega_\beta}{\Energy_\beta} Q_\beta^\ast |a(z, t)|^2.
\end{align}
The acoustic equations of motion can be formally solved using Green functions:
\begin{align}
  \tilde b_\beta(z, t) = & - \frac{\imag Q_\beta^\ast}{\Energy_\beta} 
  \int_0^\infty \total t' \ 
  |a(z-v_\beta t', t-t')|^2
  \mathrm{e}^{-(\gamma_\beta + \imag \Omega_\beta - \imag v^{(1)}_\beta q_0) t'}.
\end{align}
This allows us to combine the Raman kernel with the Brillouin response in a single
function $r_\text{eff}(z, t)$:
\begin{align}
  \left[ \gamma_\alpha + \partial_t + v_\alpha^{(1)} \partial_z 
  + \imag \frac{v_\alpha^{(2)}}{2} \partial_z^2 \right] a(z, t)
  = &
  \imag a(z, t) \int_0^\infty \total t' \ \int_0^\infty \total z' \ 
  r_\text{eff}(z', t') |a(z-z', t-t')|^2,
\end{align}
\end{widetext}
where the effective response function is spatially nonlocal due to the 
travelling nature of the sound waves in addition to the temporal non-locality 
that is known from Raman scattering:
\begin{align}
  r_\text{eff}(z, t) = & \delta(z) r(t) - 
  \sum_\beta
  \frac{2 \omega_0 \Omega_\beta |Q_\beta|^2 }{\Energy_\alpha \Energy_\beta} 
  \mathrm{e}^{-\gamma_\beta t}
  \\
  \nonumber
  & \quad \times
  \sin\Big[\Big(\Omega_\beta - v^{(1)}_\beta q_0\Big) t\Big] \  
  \delta\Big(z - v^{(1)}_\beta t\Big) .
\end{align}
This spatial nonlocality is the main difference between FBS and Raman 
scattering.
This nonlocality would be relevant for FBS involving a Dirac cone in the 
acoustic dispersion relation at $q=0$.
However, in practice FBS occurs at acoustic band edges, where the
the acoustic group velocity $v^{(1)}_\beta$ so small that the phonon propagation 
length (\ie the spatial nonlocality) is but a few nanometers.
In this case, spatial nonlocality is negligible and intra-mode forward 
Brillouin scattering can be regarded as a simple contribution to the 
low-frequency tail of the Raman spectrum:
\begin{align}
  r_\text{eff}(t) = & r(t) - \sum_\beta 
  \frac{2 \omega_0 \Omega_\beta |Q_\beta|^2 }{\Energy_\alpha \Energy_\beta} 
  \mathrm{e}^{-\gamma_\beta t}
  \sin(\Omega_\beta t).
\end{align}
FBS can therefore be expected to contribute to the formation and the 
self-frequency shift of nanosecond-scale solitons in 
waveguides~\cite{Gordon1986}.

\section{Discussion}
\label{sec:discussion}

After the analysis of the previous sections we now combine the results 
obtained in the different regimes and discuss their implications.

%%%%%%%%%%%%%%%%%%%%%%%%%%%%%%%%%%%%%%%%%%%%%%%%%%%%%%%%%%%%%%%%%%%%%%%%%%%%%%%%
\subsection{Phase modulation versus intensity modulation}

We found in Sec.~\ref{sec:dispersionless}, that intra-mode FBS without optical 
dispersion has a closed analytical solution in the form of a simple integral.
One key observation within this result is that the acoustic field does not
affect the optical intensity, \ie neither the optical beat pattern nor the 
acoustic amplitude vary along the waveguide.
Consequently, dispersionless intra-mode FBS results only in a phase modulation 
of the optical signal.
Thus, the acoustic amplitude is constant throughout the waveguide and the
phase modulation grows linearly.

The situation changes fundamentally if the waveguide is dispersive.
The optical dispersion slowly shifts the various (anti-)Stokes orders out of
phase and thereby creates an FBS-related beating pattern, \ie an optical 
intensity modulation.
Depending on the sign of the dispersion constant, this new beating pattern
can have the opposite or the same sign as the intensity fluctuation that
drives the acoustic wave in the first place.
In the former case, the total optical intensity variations (and thus the 
acoustic amplitude) diminish along the waveguide.
However, the entropy of the fluctuating pump cannot disappear, so the reduction
in the intensity modulation has to be countered by a phase modulation.
In the other case (additive beat patterns), optical intensity fluctuations (as well
as the acoustic amplitude) are amplified along the waveguide and both the
intensity and the phase modulation grow exponentially along the waveguide.
We note that, having used a perturbative approach to solve this problem, 
these results are strictly only valid for weak dispersion and weak intensity 
contrasts --- this approximation will therefore break down as the Stokes grows 
to be comparable with the pump. 
Nevertheless these results can be expected to remain qualitatively valid even
in the cases of strong dispersion and large intensity contrasts.

%%%%%%%%%%%%%%%%%%%%%%%%%%%%%%%%%%%%%%%%%%%%%%%%%%%%%%%%%%%%%%%%%%%%%%%%%%%%%%%%
\subsection{Measurement of FBS}
\label{sec:measurement}

FBS occurs in various fields such as photonic crystal fibres or integrated 
silicon waveguides.
These systems are typically studied with different experimental techniques
because of the very dissimilar loss regimes, power handling and waveguide
lengths.
Most fiber experiments capture the modulation of the optical wave with high
fidelity either by using an interferometric setup or by measuring in another 
polarization.
They usually do not inject a dedicated Stokes seed and instead observe 
scattering from phonons that are either thermally excited or generated by
the intensity noise of the pump laser.
The long interaction lengths and power handling in fiber optics allows for 
complex interplay of FBS with other nonlinearities and often result in cascaded 
dynamics. 
Extreme examples of this are lasers that are passively mode-locked at the 
acoustic resonance frequency~\cite{Kang2013,Stiller2013}.
It should be stressed that in the case of a straight dispersionless waveguide
and for strict intra-mode scattering the intensity variations at the end of a 
waveguide do not differ from those at the front, because the beat of the pump
and Stokes lines and among the Stokes lines vanishes exactly.
This is the reason why many measurements 
are either done in orthogonal polarizations or in interferometric setups that
are designed to eliminate the main pump 
line~\cite{Elser2006,Dainese2006,Beugnot2007,Kang2008,Stiller2011,Zhong2015}.
Through the amplitude and phase modifications of the individual Stokes order 
this leads to a beat signal, which can then be easily detected by a 
photodiode and analyzed in the RF domain.
It should be noted that the relative amplitudes of the lines in the RF spectrum
can differ considerably from relative amplitudes in the actual optical 
spectrum.

In contrast, experiments on FBS in integrated photonics often require a 
pump-probe configuration with fairly strong Stokes seeds in order to lift the 
spectral peaks above the detection threshold.
As a consequence, the measurements are usually not sensitive enough to
distinguish between ``stimulated'' (in the sense of ``self-amplifying'') 
dynamics and an ``optically driven'' phase modulation.
Furthermore, these experiments may not require special ``phase-mixing'' 
elements (\eg interferometric loops) to convert the pure phase modulation into 
a partial intensity modulatoin, because (narrow-band) grating couplers 
readily introduce phases and the waveguides are dispersive in themselves.
Quite often, the interaction lengths are too short for cascading to set in.
Without the need to capture cascading, the theoretical description is 
often restricted to basic coupled-mode theory borrowed from backward SBS.
Only recently, an extension of the coupled-mode formulation to include the
first anti-Stokes line has led to the realization~\cite{Rakich2016} that 
self-amplifying FBS is not generally expected in integrated waveguides.
Care should be taken when studying FBS in weakly dispersive integrated 
waveguides with broad-band couplers (or even butt-coupling):
without the phase modification to the individual Stokes orders, this will
only result in a pure phase modulation that will go unnoticed if directly 
analyzed with a photo diode.

%%%%%%%%%%%%%%%%%%%%%%%%%%%%%%%%%%%%%%%%%%%%%%%%%%%%%%%%%%%%%%%%%%%%%%%%%%%%%%%%
\subsection{Stimulated versus Raman-like scattering}

\begin{figure}
  \includegraphics[width=0.43\columnwidth]{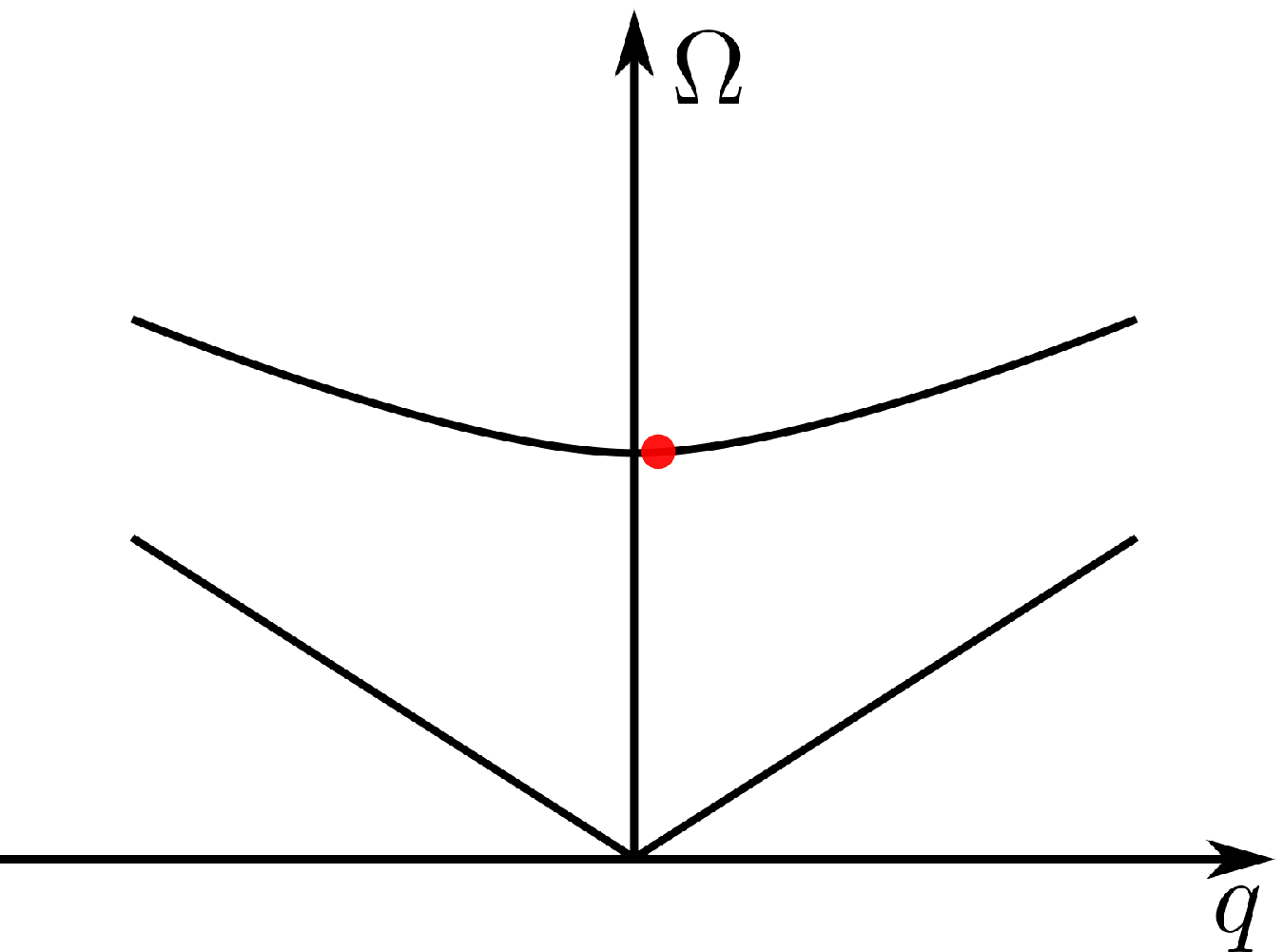}
  \hspace{0.08\columnwidth}
  \includegraphics[width=0.43\columnwidth]{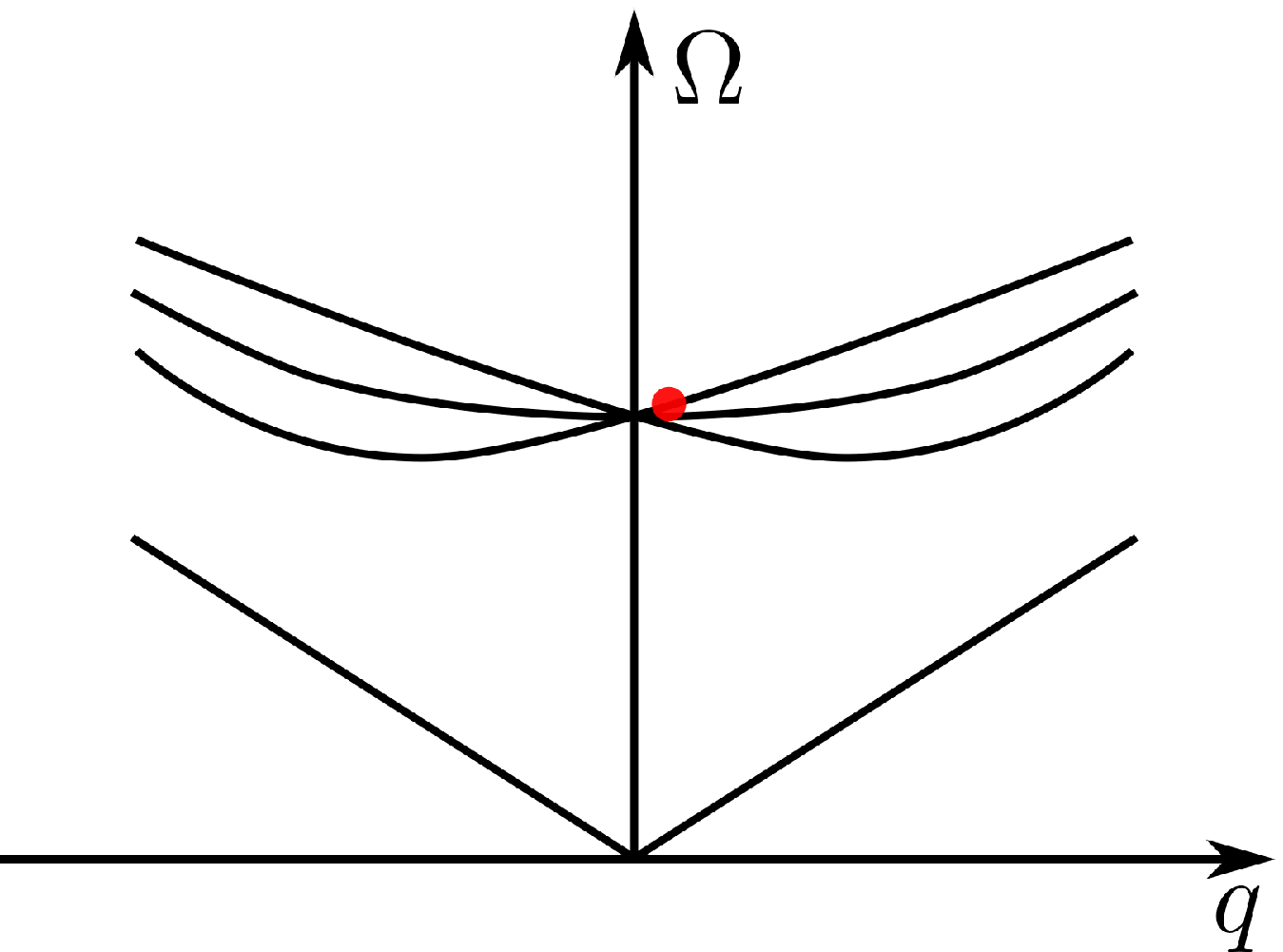}
  \caption{
    Schematic examples for acoustic dispersion relations with low acoustic 
    group velocity of the higher-order acoustic mode (red dot in left panel) 
    and with a Dirac point at $q=0$ leading to high acoustic group velocities 
    of some of the higher-order acoustic modes (red dot in right panel).
    FBS in the conventional case with low acoustic group velocity can be 
    described as a low-frequency contribution to the local Raman response.
    In contrast, FBS in the right-hand case manifests as a non-local effect 
    along the waveguide.
    It should be noted that the question whether self-amplification can occur
    is not related to this, but determined by the \emph{optical} dispersion 
    relation.
  }
  \label{fig:dirac_point}
\end{figure}

As we have mentioned in the introduction, FBS is known under different names, 
especially as ``Raman-like'' or as ``forward stimulated'' scattering.
Each of them is justified, because they indicate features that may or may
not be dominant depending on the specific type of waveguide and the measurement
technique employed.
For example, in Sec.~\ref{sec:raman-like}, we have clearly shown that intra-mode
FBS is phenomenologically indistinguishable from a low-frequency tail of the 
Raman spectrum if and only if the acoustic group velocity (\ie the velocity at
which an acoustic pulse travels along the waveguide) is negligibly small.
This is usually the case in intra-mode FBS and inter-mode FBS between degenerate
optical modes (see Fig.~\ref{fig:dirac_point}).
In these systems the term ``Raman-like'' is well justified.
Large acoustic group velocities can occur for inter-mode FBS or if the acoustic
dispersion relation has a Dirac point at $q=0$. 
The formation of Dirac points through the combination of symmetry-related and
accidental degeneracies at the $\Gamma$-point is well understood in the context 
of photonic crystals~\cite{Huang2011} and similar effects are expected in 
mechanical systems, as well.
In these two cases, FBS differs from Raman scattering in that it becomes 
non-local over the length scale of the acoustic decay length 
(approx. $50-100\,\mu$m), which can be very relevant in integrated 
photonics~\cite{Wolff2015b}.

Both Brillouin scattering processes, forward and backward, are referred to as 
\emph{Stimulated} Brillouin Scattering (SBS) in parts of the literature, while
some authors restrict this terminology to the backward scattering process, only.
The observation that FBS in the absence of dispersion does not result in any 
amplification of the Stokes signal prompts the question: 
should FBS be classified as a stimulated process?  
The answer to this depends on one's definition of what a stimulated process 
entails. 
\comment{
  A stimulated process may be defined as one in which the probability of a 
  transition between states --- in this case a phonon generated by the transition 
  between a high-energy photon state to a lower energy photon state --- is 
  increased in proportion to the population of the emitted quantum. 
  However there are also significant differences between Brillouin transitions, 
  which involve bosons, and the transitions in a conventional laser, which 
  involve fermions. 
  In particular, Brillouin scattering can be described in a purely classical 
  form, and Planck's constant does not appear in the rate equations.
  In the end, the classification of such classical systems, which include not 
  only (backward) SBS, but also Raman and X-ray lasers, as ``stimulated'' has 
  become a matter of convention and is widely accepted.
}
In the context of backward Brillouin scattering and Raman scattering, the term 
``stimulated'' scattering has been used since the early 
papers~\cite{Bloembergen1965,Tang1966} to denote the positive feedback that results in 
self-amplification of both the optical and acoustic fields. 
Using this definition a clear distinction is drawn between the spontaneous 
processes, in which the pump signal is too weak to cause a runaway effect so 
strong that thermal excitations are amplified to levels comparable to the pump 
power, and the stimulated process, in which significant self-amplification 
occurs and the backscattered Stokes signal grows exponentially along the 
length of the waveguide. 
This definition applies to FBS provided that the waveguide exhibits an 
appropriate optical dispersion, as we have shown in Sec.~\ref{sec:dispersive}, 
but does not if this condition is not met. 
Adopting this pragmatic approach, it seems fair to call FBS a stimulated 
process, but to keep in mind that whether or not the effect is ``stimulated'' 
will always depend on the details of the waveguide and experiment. 
While the presence of dispersion certainly allows forward amplification, our 
analysis shows that for the zero dispersion case, or more physically, for 
distances shorter than the dispersion length, there is no significant 
self-amplification at powers where amplification would certainly be seen in 
the backwards case. 
This suggests that while FSBS is a defensible terminology, perhaps FBS is to 
be preferred in generality.

%%%%%%%%%%%%%%%%%%%%%%%%%%%%%%%%%%%%%%%%%%%%%%%%%%%%%%%%%%%%%%%%%%%%%%%%%%%%%%%%
\subsection{Similarities to cavity opto-mechanics and discrete diffraction }

The problem of FBS bears similarities to other fields of optics apart from 
backward SBS and Raman scattering.
One example is the close similarity between FBS in dispersive waveguides and
side-band heating and cooling in cavity opto-acoustics~\cite{Bahl2012}.
In both cases the opto-acoustic interaction can either suppress acoustic
vibrations or lead to an exponential growth.
The particular outcome is determined by the imbalance of the optical 
density of states on the red and the blue side of the optical pump beam.
Such a connection is somewhat expected given the intimate connection between
both fields~\cite{VanLaer2016}.
However, there is one fundamental difference:
in cavity opto-acoustics, the acoustic oscillations are amplified or suppressed
over time, whereas in FBS they are modified along the waveguide.

While cavity opto-acoustics and FBS are clearly based on the same physics, the
intimate connection between FBS and discrete diffraction in waveguide arrays 
is on a formal rather than a physical level.
The various Stokes orders constitute channels of light propagation similar to
an array of waveguides and a transversal acoustic mode provides a coupling 
between these channels, whose magnitude is proportional to the acoustic 
amplitude.
It is a fortunate coincidence that in the absence of optical dispersion the 
acoustic amplitude is constant along the waveguide, \ie that the process of FBS
is formally equivalent to an array of parallel waveguides.
As a result, the evolution of the optical intensity through the Stokes orders
along the waveguide is governed by the same equations and the solution of the
two-tone problem (Sec.~\ref{sec:two-tone}) is identical to the diffraction of 
light in an array of parallel waveguides.
One of the intriguing features of this analogy is that the inter-channel 
coupling coefficient (or alternatively the effective length $\zeta$ of the
waveguide array) can be tuned with extreme flexibility via the optical pump
power.
This similarity raises the possibility for FBS-based quantum-walk experiments.

%%%%%%%%%%%%%%%%%%%%%%%%%%%%%%%%%%%%%%%%%%%%%%%%%%%%%%%%%%%%%%%%%%%%%%%%%%%%%%%%
\subsection{Summary and conclusions}

We have presented here a new and rigorous formulation of the forward Brillouin 
scattering process, which implicitly includes all coupling amongst the comb of 
cascaded orders. 
Because the differences between the quantum and classical formulations are 
minimal, we have chosen the more general case for our derivation of the 
equations of motion, beginning with the full optomechanical Hamiltonian of the 
system. 
This formalism can therefore be applied to problems of interest in the limit of 
low photon/phonon numbers, and would be especially useful for the prediction and 
interpretation of single phonon scattering experiments.

The established description of FSB is derived from the theory of the backward 
SBS process in fibres. 
In this, the pump and Stokes fields are modelled as independent, spectrally 
extremely narrow fields.
This is partially motivated by the strong dependence of the acoustic frequency 
on the optical wave length and the narrow acoustic resonances. 
Cascading effects can occur in backward SBS~\cite{Buettner2014a}, but are not a 
very common feature. 
Instead, exponential growth of the optical and acoustic amplitudes along the 
waveguide is the most common manifestation.
In contrast, FBS is mostly independent of the optical wavelength, it has a
strong tendency towards cascading and requires the help of optical dispersion 
to have any impact on the optical and acoustic intensities. 
Thus, the multi-mode ansatz that is so successful in BSBS is not ideal for FBS. 
We find that a description coupling only one broad-band optical field to an 
acoustic field is much better suited.

In particular for the dispersionless case we find closed-form solutions, in 
which the amplitude of the cascaded orders is given simply by a series of 
Bessel functions. 
The simplicity of this result follows from one of the main insights arising 
from this formalism: that the acoustic field exerts, in the absence of 
dispersion, a pure phase modulation on the optical field. 
This implies that for dispersionless waveguides FBS is not capable of 
amplifying the intensity beat between a pump and a Stokes signal.

%%%%%%%%%%%%%%%%%%%%%%%%%%%%%%%%%%%%%%%%%%%%%%%%%%%%%%%%%%%%%%%%%%%%%%%%%%%%%%%%
%%%%%%%%%%%%%%%%%%%%%%%%%%%%%%%%%%%%%%%%%%%%%%%%%%%%%%%%%%%%%%%%%%%%%%%%%%%%%%%%
\section{Acknowledgements}

We acknowledge financial support of the Australian Research
Council via the ARC Center of Excellence CUDOS (CE110001018) and
its Laureate Fellowship (Prof. Eggleton, FL120100029) program.
Furthermore, we are deeply indebted to Dr Mark Craddock for insightful
discussions.

\begin{appendix}

\section{Derivation of first-order perturbative dispersion terms}
\label{appx:perturbation}
Here we derive the corrections to the optical and acoustic fields
for the dispersive dynamics shown in Eqs.~\eqref{eq:h1corr}
and~\eqref{eq:g1corr}.

We start with the dispersion-free stationary two-tone solution:
\begin{align}
  h_0(z, t) \approx & \log | F_0 |,
  \\
  g_0(z, t) = & \imag G_0 \exp\Big(- \imag \Omega_0 \frac{vt - z}{v} \Big),
  \\
  j_0(z, t) = & \frac{2 G_0 \omega_0 z}{v} 
  \sin\Big(\Omega_0 \frac{vt - z}{v} \Big),
\end{align}
with real positive $G_0$.
The correction to $p$ satisfies:
\begin{align}
  \partial_t h_1 + [v + 2 w (\partial_z j_0)] \partial_z p = & 
  - w \partial_z^2 j_0,
\end{align}
with the boundary condition at $z=0$
\begin{align}
  h_1(0, t) = 0.
\end{align}
The derivatives of the unperturbed phase function are:
\begin{align}
  \partial_z j_0 = & \frac{G_0\omega_0}{v} \sin\Big(\Omega_0 \frac{vt - z}{v}\Big)
  - \frac{G_0 \omega_0 \Omega_0 z}{v^2} \cos\Big(\Omega_0 \frac{vt - z}{v}\Big),
  \\
  \partial_z^2 j_0 = & -\frac{4G_0\omega_0\Omega_0}{v^2} \cos\Big( \Omega_0 \frac{vt - z}{v} \Big)
  \\
  \nonumber
  & \quad - \frac{2G_0 \omega_0 \Omega_0^2 z}{v^3} \sin\Big(\Omega_0\frac{vt-z}{v}\Big).
\end{align}
The dispersive correction to the effective group velocity is of order $w$ and 
therefore introduces corrections of order $w$.
We may therefore approximate:
\begin{align}
  \nonumber
  (\partial_t + v \partial_z) h_1 = & 
  \frac{2 w \omega_0 \Omega_0 G_0}{v^3} 
  \Big[ 2 v \cos\Big( \Omega_0 \frac{vt - z}{v} \Big)
  \\
  & \quad 
  + \Omega_0 z \sin\Big(\Omega_0\frac{vt-z}{v}\Big)\Big].
\end{align}
Through the substitution
\begin{align}
  y = & vt + z, & s = & t - \frac{z}{v};
  \\
  z = & \frac{y - vs}{2}, & t = & \frac{y + vs}{2v},
\end{align}
this can be reduced to the ordinary differential equation 
\begin{align}
  v \partial_y = & 
  \frac{2 w \omega_0 \Omega_0 G_0}{v^3} 
  \Big[ 2 v \cos \Omega_0 s 
  + \frac{\Omega_0 (y - vs)}{2} \sin\Omega_0 s\Big].
\end{align}
with the explicit 
solution:
\begin{widetext}
\begin{align}
  h_1(y, s) = & \frac{w \omega_0 \Omega_0 G_0}{v^4} \int_{0}^{y}
  \total y' \ 
  \Big[ 2 v \cos (\Omega_0 s) 
%    \\
%  \nonumber & \quad 
  + \Omega_0 \frac{y' - vs}{2} \sin(\Omega_0 s) \Big] + C(s),
    \\
  = & \frac{w \omega_0 \Omega_0 G_0}{v^4} \Big[ 2 v y \cos (\Omega_0 s) 
%    \\
%    \nonumber & \quad
  + \frac{\Omega_0}{4} (y^2 - 2 y vs) \sin(\Omega_0 s) \Big] + C(s),
\end{align}
where $C(s)$ is defined through the boundary condition expressed in the 
transformed coordinates \mbox{$h_1(vs, s) = 0$}.
Therefore, the first order correction to $p$ is:
\begin{align}
  h_1(y, s) = & \frac{w \omega_0 \Omega_0 G_0}{v^4} \Big[
    2 v(y - vs) \cos(\Omega_0 s) 
%    \\
%    \nonumber \quad &
    + \frac{\Omega_0}{4} (y - vs)^2 \sin(\Omega_0 s) 
  \Big];
  \\
  h_1(z, t) = & \frac{w \omega_0 \Omega_0 G_0}{v^4} \Big[
  4 vz \cos\Big(\Omega_0 \frac{vt-z}{v}\Big) 
%    \\
%    \nonumber \quad &
    + \Omega_0 z^2 \sin\Big(\Omega_0 \frac{vt-z}{v}\Big) 
  \Big].
\end{align}
This recovers Eq.~\eqref{eq:h1corr} in the main text.

The remaining step is to find the correction to the acoustic problem:
\begin{align}
  g_0(z, t) + g_1(z, t) = & \imag \Omega_0 \int_0^\infty \total t' \ 
  \text{e}^{-\gamma t' - \imag \Omega_0 t'} \big|A(t-t'-z/v)\big|^2 \exp[2 h_1(z, t-t')]
  \\
  \approx & \imag \Omega_0 \int_0^\infty \total t' \ 
  \text{e}^{-\gamma t' - \imag \Omega_0 t'} \big|A(t-t'-z/v)\big|^2 [1 + 2 h_1(z, t-t')]
  \\
  = & g_0(z, t) + \underbrace{2 \Omega_0 \imag \int_0^\infty \total t' \ 
  \text{e}^{-\gamma t' - \imag \Omega_0 t'} \big|A(t-t'-z/v)\big|^2 h_1(z,t-t')}_{=g_1(z, t)}.
\end{align}
Assuming that the intensity of the incident optical signal varies only weakly
(weak incident Stokes seed, \eg due to thermal fluctuations), we can replace 
\mbox{$|F(t)|^2 \approx |F_0|^2$} in the second term:
\begin{align}
  g_1(z, t) \approx & \frac{2 \imag w \omega_0 \Omega_0^2 G_0 |F_0|^2}{v^4} 
  \int_0^\infty \total t' \ \text{e}^{-\gamma t' - \imag \Omega_0 t'} 
  \Big[ 4 vz \cos\Big(\Omega_0 \frac{vt-vt'-z}{v}\Big) 
    + \Omega_0 z^2 \sin\Big(\Omega_0 \frac{vt-vt'-z}{v}\Big) 
  \Big]
  \\
  \approx &
  \frac{2 \imag w \omega_0 \Omega_0^2 G_0 |F_0|^2}{v^4 \gamma} 
  \Big[ 4 v z + \imag \Omega_0 z^2 \Big] \exp\Big(\imag \Omega_0 \frac{z - vt}{v}\Big)
\end{align}
where only retained the resonant contribution to the convolution.
This is the second expression required in the main text.
\end{widetext}

\end{appendix}

\end{document}